\documentclass[preprint,11pt]{elsarticle}
\usepackage[left=2.5cm,right=2.5cm,top=2cm,bottom=3cm]{geometry}
\usepackage{graphicx} 
\usepackage{siunitx}
\usepackage{placeins}
\usepackage{arydshln}
\usepackage{amsmath}
\usepackage{tabularx}
\usepackage{booktabs}
\usepackage{array}
\usepackage{graphicx}
\usepackage{subcaption}
\usepackage{multirow}
\usepackage{booktabs}   
\usepackage{siunitx}    
\usepackage{multirow}
\usepackage{threeparttable}
\usepackage{tabularx} 
\usepackage{booktabs} 
\usepackage{amsfonts}
\usepackage{subcaption}

\usepackage{hyperref}
\hypersetup{
    colorlinks=true,
    linkcolor=blue,
    filecolor=magenta,      
    urlcolor=cyan,
    citecolor=blue,
}

\newcolumntype{C}[1]{>{\centering\let\newline\\\arraybackslash\hspace{0pt}}m{#1}}

\biboptions{numbers,sort&compress}

\journal{Energy Conversion and Management}

\begin{document}

\begin{frontmatter}

\date{}

\title{Geographic variability in reanalysis wind speed biases: A high-resolution bias correction approach for UK wind energy}

\author[inst1]{Yan Wang\corref{cor1}}
\ead{yan.wang22@imperial.ac.uk}

\author[inst1]{Simon C. Warder}
\author[inst1]{Ellyess F. Benmoufok}
\author[inst2]{Andrew Wynn}
\author[inst2]{Oliver R. H. Buxton}
\author[inst3]{Iain Staffell}
\author[inst1]{Matthew D. Piggott}

\affiliation[inst1]{organization={Department of Earth Science and Engineering},
            addressline={Imperial College London}, 
            city={London},
            postcode={SW7 2AZ}, 
            country={UK}}
\affiliation[inst2]{organization={Department of Aeronautics},
            addressline={Imperial College London}, 
            city={London},
            postcode={SW7 2AZ}, 
            country={UK}}
\affiliation[inst3]{organization={Centre for Environmental Policy},
            addressline={Imperial College London}, 
            city={London},
            postcode={SW7 1NE}, 
            country={UK}}

\cortext[cor1]{Corresponding author.}

\begin{abstract}
Reanalysis datasets have become indispensable tools for wind resource assessment and wind power simulation, offering long-term and spatially continuous wind fields across large regions. However, they inherently contain systematic wind speed biases arising from various factors, including simplified physical parameterizations, observational uncertainties, and limited spatial resolution. Among these, low spatial resolution poses a particular challenge for capturing local variability accurately. Whereas prevailing industry practice generally relies on either no bias correction or coarse, nationally uniform adjustments, we extend and thoroughly analyse a recently proposed spatially resolved, cluster-based bias correction framework. This approach is designed to better account for local heterogeneity and is applied to 319 wind farms across the United Kingdom to evaluate its effectiveness. Results show that this method reduced monthly wind power simulation errors by more than 32\% compared to the uncorrected ERA5 reanalysis dataset. The method is further applied to the MERRA-2 dataset for comparative evaluation, demonstrating its effectiveness and robustness for different reanalysis products. In contrast to prior studies, which rarely quantify the influence of topography on reanalysis biases, this research presents a detailed spatial mapping of bias correction factors across the UK. The analysis reveals that for wind energy applications, ERA5 wind speed errors exhibit strong spatial variability, with the most significant underestimations in the Scottish Highlands and mountainous areas of Wales. These findings highlight the importance of explicitly accounting for geographic variability when correcting reanalysis wind speeds, and provide new insights into region-specific bias patterns relevant for high-resolution wind energy modelling.
\end{abstract}


\begin{highlights}
    \item Spatially resolved bias correction improves wind power modelling in the UK.
    \item Applied the method to ERA5 and MERRA-2 reanalysis datasets for performance comparison.
    \item Reduced monthly wind power simulation errors for both reanalysis products.
    \item Spatially mapped wind speed bias correction factors across the UK.
    \item Analysed correction factor correlations with topography and turbine concentration.

\end{highlights}


\end{frontmatter}




\section{Introduction}
Wind energy plays a central role in the global transition towards sustainable energy and continues to rank among the fastest-growing and most cost-competitive renewable technologies \cite{warder2025mapping}. By the end of 2024, global installed wind power capacity surpassed 1130 GW, with a record 116 GW added in that year alone \cite{WWEA2025}.
Wind energy made a substantial contribution to global renewable energy expansion in 2024 and is projected to remain a cornerstone of the global renewable energy portfolio \cite{IRENA2024}. However, fully realizing the potential of wind power entails overcoming challenges that span turbine-level aerodynamics, their interactions with atmospheric flows, and the coordination of plant-scale control systems \cite{bempedelis2023data}. These challenges are likely to intensify as wind energy penetrates deeper into national electricity grids \cite{drew2013estimating,veers2019grand}. Effective modelling strategies are therefore essential to support informed decision-making by diverse stakeholders, including government agencies, energy providers, and, in the UK specifically, organizations such as the Crown Estate \cite{bilas2022supporting,o2012investigation,simcock2016procedural}.

 Accurate wind resource assessment is a prerequisite for reliable wind power modelling and effective integration into electricity grids \cite{clare2023unsupervised}. In practice, reliable wind energy simulations rely heavily on high-quality meteorological data. Among the various sources available, reanalysis datasets \cite{doddy2021reanalysis}, such as ERA5 and MERRA-2, have emerged as essential products in both academic studies and industry applications related to wind energy. Reanalysis datasets are a fundamental component in modern atmospheric and renewable energy research. These datasets combine historical observational data with modern numerical weather prediction models to generate long-term, spatially continuous, and meteorologically consistent atmospheric records \cite{gelaro2017modern,compo2011twentieth}. They represent a comprehensive effort to reconstruct past atmospheric conditions by assimilating a vast array of observational data into a consistent numerical weather prediction (NWP) model using data assimilation techniques \cite{fujiwara2016introduction,al2010review,hersbach2018operational}.
 
Despite their many advantages, reanalysis datasets exhibit several important limitations that can affect the reliability of wind assessments if left unaddressed \cite{wang2020evaluation,parker2016reanalyses}. One of the most critical limitations is the presence of systematic wind speed biases \cite{cucchi2020wfde5}, particularly in areas with complex terrain or near coastlines, due to their limited spatial resolution. 
It has been shown that the bias increases with both site elevation and measurement height, leading to systematic underestimation at higher altitudes \cite{rose2016quantifying}. Over France, ERA5 was found to outperform other reanalyses but still systematically underestimates wind speeds in mountainous regions \cite{jourdier2020evaluation}. Similar patterns were confirmed in global tall-tower comparisons, where ERA5 performed reliably in flat and offshore sites but showed large biases in mountainous areas due to its coarse resolution \cite{gualtieri2021reliability,potisomporn2023era5}. In Denmark, evaluations showed a systematic overestimation of inland wind speeds by ERA5, in the context of a relatively flat topography \cite{benmoufok2024improving}. Multiple studies indicate that reanalysis datasets, particularly ERA5, achieve higher accuracy over flat terrain and offshore settings, with offshore performance generally exceeding onshore performance, while errors increase markedly in complex topography \cite{potisomporn2023era5,gualtieri2022analysing,alkhalidi2025evaluating}. One objective of this study is to provide a comprehensive assessment of the spatial distribution of ERA5 wind-speed biases across the United Kingdom and to investigate their relationship with topographic characteristics.

Many recent studies have explored wind bias correction techniques, evidencing their broad applicability across environmental, climate, and meteorological modelling domains. For instance, a fetch-based statistical correction was developed for NWP winds over small inland water bodies, which nearly eliminated mean bias and enabled downscaling of wind fields to 500 m resolution \cite{holman2017fetch}. In the context of tropical cyclones, a multivariate regression adjustment was applied to ECMWF ensemble forecasts, significantly improving the accuracy of hurricane intensity and wind-structure predictions, particularly in peak wind speeds \cite{aijaz2019bias}. Climate simulations have likewise benefited: multiple bias-correction methods were tested on regional climate model wind outputs, showing that all tested approaches effectively reduced systematic errors, while distribution-driven corrections achieved the greatest improvements in the statistical representation of wind speed distributions \cite{li2019statistical}. More recently, advanced machine learning techniques have also been employed. For example, a deep learning framework was developed to perform real-time bias correction on ECMWF wind-field forecasts, achieving substantial reductions in systematic wind speed and direction errors \cite{zhang2023deep}. These efforts demonstrate that wind bias correction can benefit a wide range of applications, spanning hurricane forecasting, urban wind-field mapping, and climate model projections.

While bias correction methods have been widely applied in meteorological and climate-related contexts, recent years have also seen growing efforts to improve the accuracy of wind power simulations. A wide range of techniques have been developed to address discrepancies between reanalysis-driven wind power simulations and observed generation data. One of the most widely adopted approaches is linear scaling, which has been implemented at the national level to adjust ERA-Interim wind speeds, achieving high accuracy in simulating aggregated fleet performance across Europe \cite{staffell2016using,staffell2016renewables}. In the contiguous U.S., a regression-based correction framework reduced ERA5 wind power biases (including seasonal extremes) by up to 20\% \cite{wilczak2024evaluation}. Beyond these linear methods, more advanced statistical approaches have also been explored. For example, a non-Gaussian bias correction framework was applied to assess the reliability of wind power operations under climate change, enabling a more accurate representation of wind variability and extremes in power system simulations \cite{zhang2021assessing}. In parallel, machine learning models have been developed to predict correction factors, showing a strong ability to capture nonlinearity and complex interactions \cite{hallgren2023machine}. Bias correction has also improved long-term wind energy assessments, such as in the Iberian Peninsula, where adjustment techniques refined future projections of offshore wind resources and increased the reliability of reanalysis-driven power estimates \cite{costoya2020using}. Together, these diverse studies highlight that accurate wind energy modelling demands flexible, data-driven correction methods tailored to local spatial and temporal contexts. Yet, most existing approaches remain at a coarse, national scale, neglecting local variability. Such oversimplification limits their effectiveness for detailed wind resource evaluation and power forecasting at the wind farm scale. Further research is therefore needed to exploit fine-resolution observational data and enable more granular bias correction.

Despite recent efforts to develop finer-resolution correction frameworks, their applicability remains constrained by limited geographical scope and the granularity of available data. For instance, a cluster-based correction method has been successfully applied in Denmark \cite{benmoufok2024improving}, demonstrating improvements in the accuracy of wind speed corrections for the ERA5 dataset. Despite these advancements, the method’s applicability remains geographically limited, having only been validated within the specific context of Danish wind turbines. Furthermore, it has exclusively focused on ERA5 data, with no assessment of whether this framework is transferable to other reanalysis products. This gap leaves the generality and robustness of the method across datasets uncertain. Another critical limitation is that such studies rarely incorporate detailed analyses of geographic factors, including terrain complexity and proximity to coastlines, to systematically identify the underlying causes of wind speed biases. Few studies have provided explicit quantification and spatial mapping of how local topographic variability influences wind resource assessments based on reanalysis data. This represents a significant knowledge gap, as understanding these relationships is essential for achieving robust and broadly applicable bias corrections.

To bridge these critical gaps, this work proposes a high-resolution bias correction framework tailored to the geographic diversity of the United Kingdom, and also accounting for its distinctive mix of onshore and offshore wind-farm sites. Specifically, this work comprises the following components:

\begin{itemize}
    \item[(i)] A cluster-based bias correction framework with high spatial resolution across UK wind farms;
    
    \item[(ii)] Cross-dataset evaluation and validation of the correction methodology using both ERA5 and MERRA-2 reanalysis products to demonstrate its robustness and general transferability;
    
    \item[(iii)] Spatial mapping of wind speed correction factors to identify regional bias patterns in the UK;
    
    \item[(iv)] Analysis of seasonal patterns in wind speed correction factors across the year;
    
    \item[(v)] Interpretation of correction factor distribution in relation to terrain heterogeneity.
\end{itemize}

The paper proceeds as follows: Section \ref{sec:2} introduces the datasets and methodology, including the reanalysis datasets used for simulations, observed wind power generation records from UK wind farms, and the proposed bias-correction method. Results are presented and discussed in Section \ref{sec:3}, followed by conclusions in Section \ref{sec:4}.

\section{Methods}

\begin{figure}[htbp]
    \centering
    \includegraphics[width=1.12\textwidth]{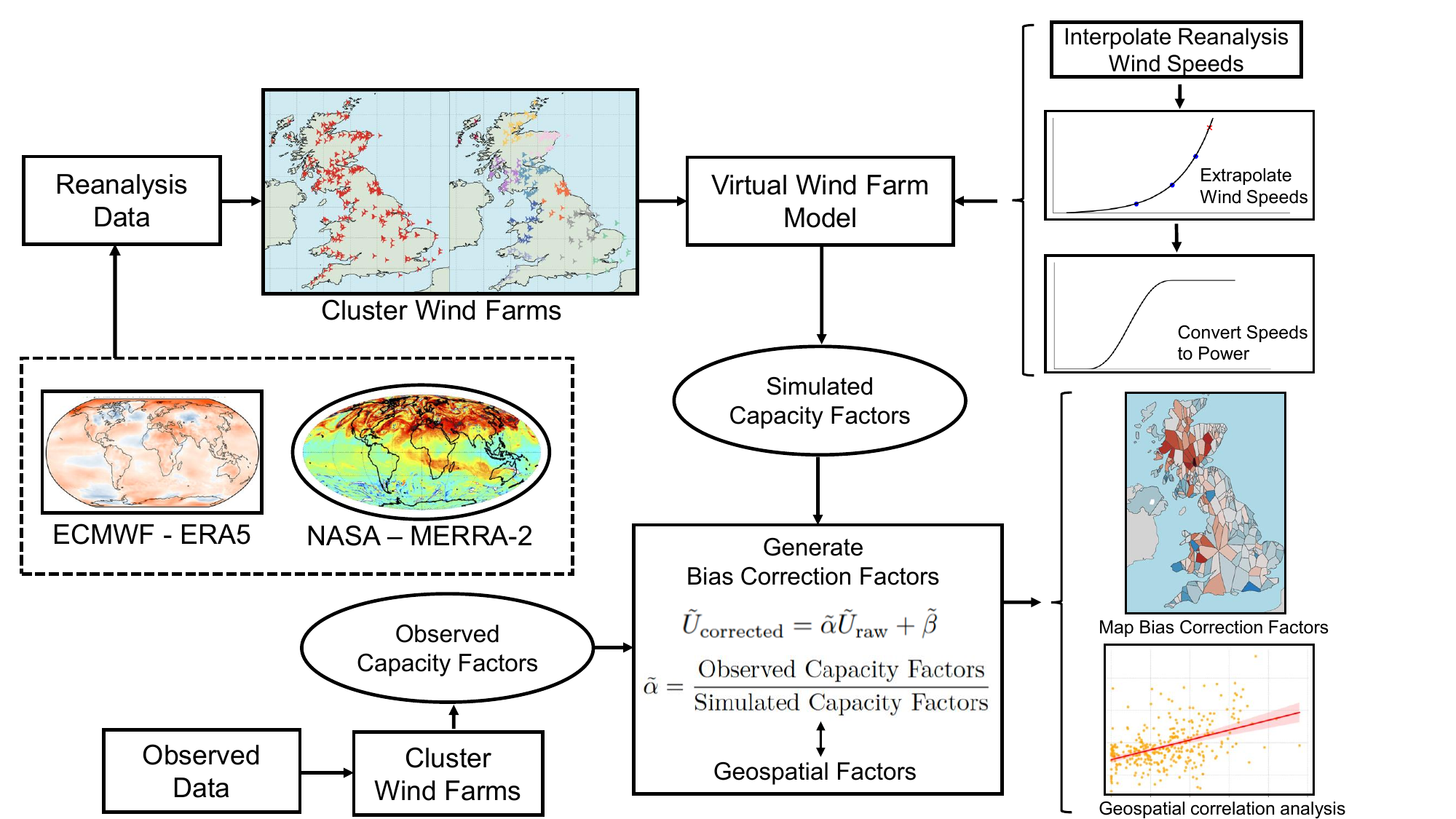}
    \caption{Overall workflow of the proposed bias correction methodology.}
    \label{fig1}
\end{figure}

The overall methodological framework proposed in this work to improve wind power simulation accuracy is illustrated in \autoref{fig1}. To evaluate and correct wind speed biases inherent in reanalysis datasets, we utilized two widely used global products: ERA5 \cite{hersbach2020era5} from the European Centre for Medium-Range Weather Forecasts (ECMWF) and MERRA-2 \cite{gelaro2017modern} from the National Aeronautics and Space Administration (NASA). Actual power generation data from wind farms served as reference observations. Each wind farm in the United Kingdom was first assigned to a spatial cluster based on geographic proximity. These clustered farms were then simulated using the Virtual Wind Farm (VWF) model \cite{staffell2016using}, which processes reanalysis wind speeds through spatial interpolation and vertical extrapolation to the turbine hub height, followed by conversion to power output using power curves, which are smoothed to represent a farm of several geographically dispersed turbines.

This modelling process produced simulated capacity factors (CFs), which were then compared against observed CFs to quantify systematic discrepancies. Although other sources of error may also contribute, we make the assumption that these discrepancies are predominantly driven by biases in the reanalysis wind speeds. Based on this assumption, wind speed correction factors were derived for each cluster using a linear formulation. Finally, we mapped the spatial distribution of these correction factors in the UK and investigated their relationships with key geospatial features. The following subsections provide detailed explanations of each component of the overall framework.

\label{sec:2}
\subsection{Reanalysis Data}
Reanalysis datasets provide long-term, spatially continuous atmospheric data by assimilating observations into numerical weather prediction models. These datasets are publicly available and cover several decades, making them a practical and cost-effective alternative to in situ measurements, especially in regions with limited observational infrastructure. 

In this study, we employ two of the most widely used global reanalysis datasets, ERA5 and MERRA-2, for wind power simulation and comparative analysis. Both datasets are known for their relatively high accuracy, comprehensive variable coverage, and suitability for long-term wind energy applications. ERA5, produced by ECMWF, is the latest generation reanalysis product from the ERA series and provides hourly data at a spatial resolution of approximately 31 km.  In contrast, MERRA-2 developed by NASA offers hourly data at a coarser spatial resolution of about 50 km.

As summarized in \autoref{tab:reanalysis_comparison}, ERA5 surpasses MERRA-2 in terms of spatial resolution, which is particularly beneficial when simulating wind conditions in heterogeneous terrains. While both datasets have been widely validated and adopted in wind energy studies, the differences between these two reanalysis datasets may lead to varying accuracy and reliability across regions. Therefore, a systematic comparison between ERA5 and MERRA-2, combined with bias correction techniques, is conducted in this study to explore their respective performance characteristics and to provide practical insights into their applicability for wind power modelling.

\newcolumntype{L}{>{\raggedright\arraybackslash}p{3.6cm}}
\newcolumntype{C}{>{\centering\arraybackslash}X}

\begin{table}[htbp]
\centering
\small
\setlength{\tabcolsep}{6pt}
\renewcommand{\arraystretch}{1.22}
\caption{Overview of two reanalysis datasets and key parameters for wind power modelling.}
\begin{tabularx}{\textwidth}{L C C}
\toprule
\textbf{Dataset Attribute} & \textbf{ERA5} & \textbf{MERRA-2} \\
\midrule
Producing Institution
& European Centre for Medium-Range Weather Forecasts (ECMWF)
& NASA Global Modeling and Assimilation Office (GMAO) \\
NWP Model
& Integrated Forecasting System
& Goddard Earth Observing System \\
Temporal Coverage
& 1940 – present\
& 1980 – present \\
Temporal Resolution
& Hourly & Hourly \\
Horizontal Resolution
& $\sim$ 31 km (0.25° lat $\times$ 0.25° lon)
& $\sim$ 50 km (0.5° lat $\times$ 0.625° lon) \\
Wind Speed Heights
& 10 m, 100 m 
& 2 m, 10 m, 50 m \\
\bottomrule
\end{tabularx}

\label{tab:reanalysis_comparison}
\end{table}

\subsection{Observed Data}
\label{sec2.2}
We assembled two independent sources of observed wind power generation to benchmark the simulations. First, we used market settlement data produced by Elexon \cite{elexon2021data}, the administrator of Great Britain’s Balancing and Settlement Code (BSC). Elexon compares contracted and actual metered volumes and publishes a comprehensive suite of operational electricity datasets at half-hourly resolution through its reporting services (formerly BMRS, now the Insights Solution \cite{elexon_bmrs}), which are widely used for market monitoring and analysis. Second, we used output inferred from Renewables Obligation Certificates (ROCs) under the UK Renewables Obligation (RO) scheme \cite{ofgem_ro,ofgem_ro_report_sy22,ofgem_ro_guidance}. The RO, introduced in 2002 \cite{ro_order_2002} and administered by Ofgem, places an obligation on licensed suppliers to present a specified number of ROCs per megawatt-hour (MWh) of electricity supplied; accredited generators receive monthly ROCs for eligible renewable output, which can be traded or retired for compliance. When converting certificates to energy, we account for technology-specific banding introduced in April 2009, which altered the number of ROCs awarded per MWh by technology (replacing the earlier one-ROC-per-MWh default) \cite{bryan2015estimating}. To enable the use of these generation records for benchmarking, we incorporated metadata on wind farm location, wind farm size, turbine type, turbine models, and turbine specifications (capacity, hub height, rotor diameter). Together, these datasets provide complementary coverage of UK renewable generation: settlement-grade market data from Elexon and policy-grade certificate records from the RO framework, offering a robust basis for analysing wind power observations in our study.

\begin{figure}[htbp]
    \centering
    \includegraphics[width=0.95\textwidth]{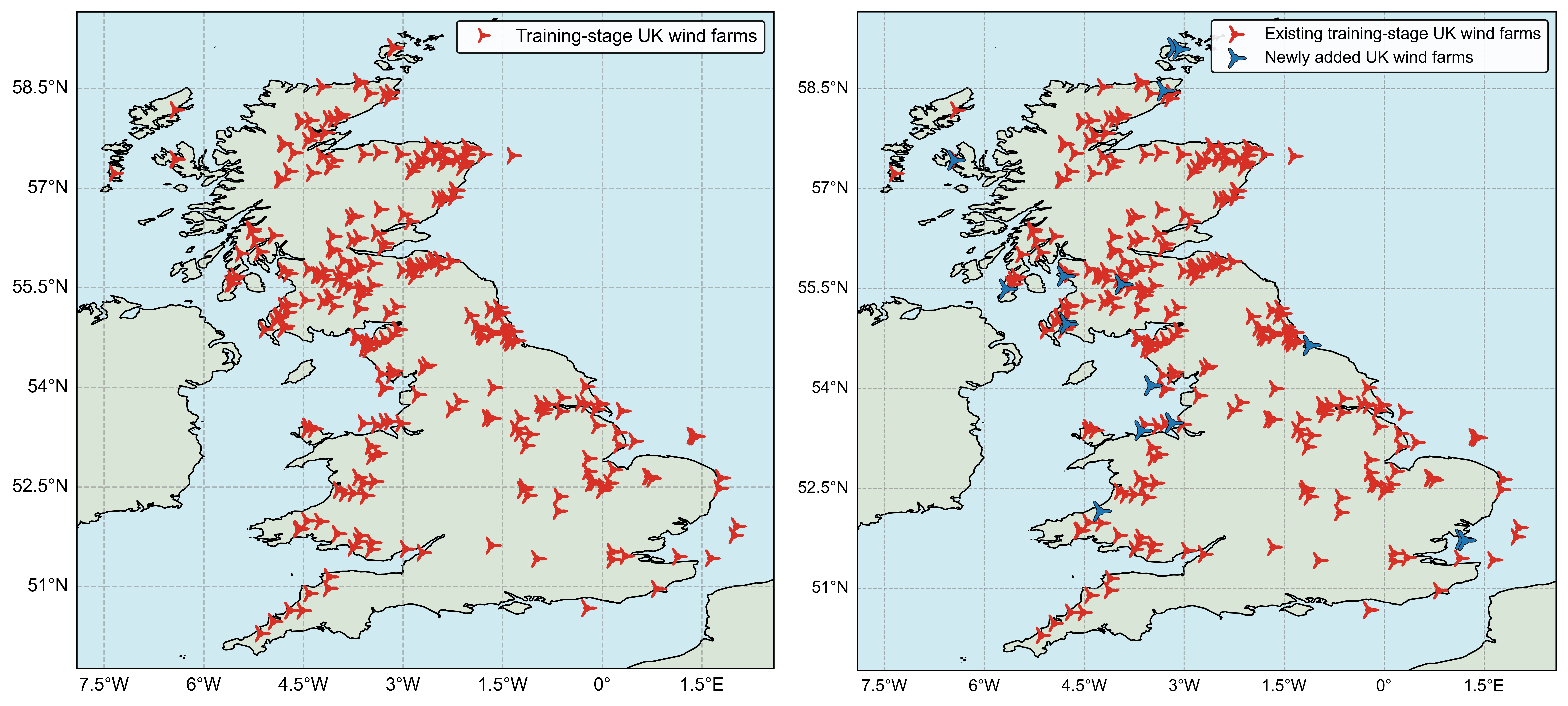}
    \caption{Distribution of UK wind farms contained within the observed data. Left: 303 wind farms used during the training phase (2015–2018, four years). Right: 319 wind farms used during the testing phase (2019, one year).}
    \label{fig2}
\end{figure}

For the purposes of model development and evaluation, we selected a four-year training period (2015–2018) and a one-year testing period (2019). During the training phase, the dataset comprised 303 wind farms and a total of 5,360 wind turbines, including 4,514 onshore and 846 offshore units. The testing dataset, covering the year 2019, consisted of 319 wind farms with 5,998 turbines, of which 4,766 were onshore and 1,232 were offshore. This split ensures that model training and testing are temporally separated, allowing for an objective evaluation of the effectiveness of the wind speed bias correction methodology. Furthermore, variations in the temporal span, along with differences in wind farm and turbine counts between the training and testing datasets, offer an opportunity to assess the model’s transferability. This capability will become increasingly important as the UK expands its onshore and offshore wind capacity to meet net-zero obligations. The distribution is shown in \autoref{fig2}, with the left panel representing the training period and the right panel representing the testing period. Red markers in \autoref{fig2} indicate wind farms in both datasets, while blue markers denote those newly added in the testing year.

\subsection{Wind Farms Clustering and Temporal Partitioning}
\label{sec2.3}

The primary objective of this study is to perform wind speed bias correction at a high spatial resolution. As a preparatory step, we refined the spatial delineation of wind farms across the United Kingdom to enable the application of location-specific correction factors. Rather than applying a single national-scale adjustment as in previous work \cite{staffell2016using}, wind farms were grouped into smaller, spatially coherent regions with similar wind conditions. This was achieved using the K-means clustering algorithm, a classical unsupervised learning method that minimizes the within-cluster sum of squared distances between samples and their assigned centroids \cite{hartigan1979algorithm,lloyd1982least,krishna1999genetic,ahmed2020k}. Formally, given feature vectors $\{ \mathbf{x}_j \}_{j=1}^{N}$ denoting the geographical coordinates of wind farm centroids and a target number of clusters $K$, 
K-means clustering solves the minimisation problem
\begin{equation}
\min_{\{C_i\}_{i=1}^K,\,\{\boldsymbol{\mu}_i\}_{i=1}^K}
\sum_{i=1}^{K} \sum_{\mathbf{x}_j \in C_i} 
\left\| \mathbf{x}_j - \boldsymbol{\mu}_i \right\|_2^{2},
\end{equation}
where $C_i$ denotes the set of samples assigned to cluster $i$ and 
$\boldsymbol{\mu}_i$ is its centroid (Euclidean mean).  The clustering was implemented in Python using scikit-learn’s \texttt{KMeans} class (\texttt{sklearn.cluster.KMeans}) \cite{pedregosa2011scikit}. We apply K-means clustering to the training-year turbines only in order to avoid information leakage, and then reuse the trained model to assign cluster labels to both training turbines and turbines that appear only in the testing year (i.e., we maintain consistent clusters across years).

\begin{figure}[htbp]
    \centering
    \includegraphics[width=1\textwidth]{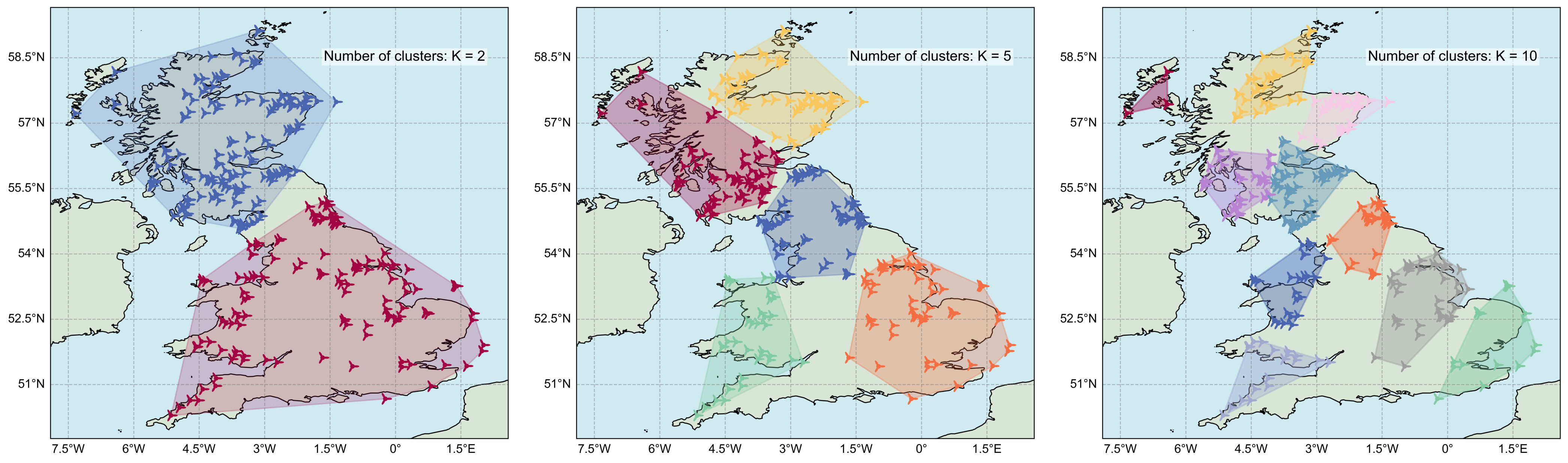}
    \caption{Multiscale K-means clustering of UK wind turbines (left to right the number of clusters $K$ = 2, 5, 10).}
    \label{fig3}
\end{figure}

\begin{figure}[htbp]
    \centering
    \includegraphics[width=1\textwidth]{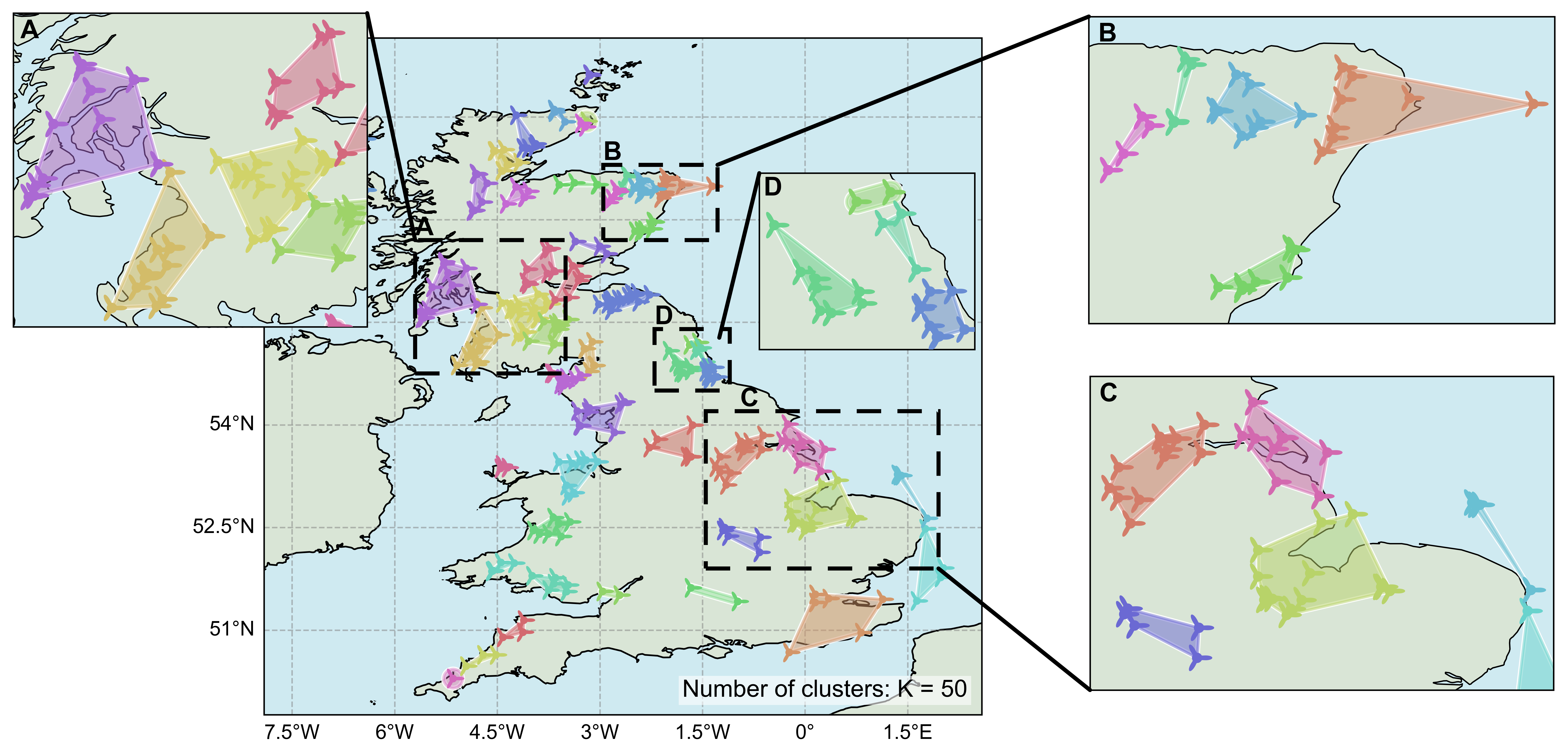}
    \caption{Fine-scale k-means clustering at $K = 50$ with zoomed-in panels A $-$ D. Insets show (A) central–western Scotland, (B) the eastern Scottish coast, (C) eastern England, and (D) Northeast England}
    \label{fig4}
\end{figure}

In our experiments, the number of clusters $K$ was varied from very coarse to fully disaggregated spatial groupings:
\[
K \in \{1,\, 2,\, 5,\, 10,\, 15,\, 20,\, 50,\, 80,\, 100,\, 120,\, 150,\, 180,\, 200,\, 250,\, 300,\, 303\},
\]
where $K=1$ corresponds to a single, national-scale correction factor (full aggregation), and $K=303$ represents fully individual correction, i.e., one correction factor per wind farm. To illustrate how cluster granularity shapes the spatial partition, \autoref{fig3} compares $K=2$ (left), $K=5$ (middle), and $K=10$ (right). With $K=2$, turbines split into a coarse national bipartition; $K=5$ yields distinct regional clusters; and $K=10$ captures finer-scale, contiguous sub-regions shaped by the turbine density field. In each panel, wind farms are enclosed by polygons, with each polygon corresponding to a single cluster. Pushing the resolution further, \autoref{fig4} uses $K=50$ to approximate wind-farm–scale segmentation. The insets highlight representative high-density regions, namely (A) central and western Scotland, (B) the eastern Scottish coast, (C) eastern England, and (D) the northeast England coastline, where compact, contiguous clusters form, enabling spatially localized bias correction and simulations. In \autoref{fig4}, clusters containing three or more wind farms are enclosed by polygons to delineate their spatial boundaries. Single-farm clusters are represented by circular envelopes, whereas two-farm clusters are visualized as narrow, elongated regions derived from the line connecting the pair of sites. 

In addition to spatial clustering, we examined different temporal granularities for applying bias correction, following a similar time-resolution differentiation approach to that proposed in \cite{benmoufok2024improving}. Specifically, wind speed data were aggregated and corrected at four time resolutions: monthly, bimonthly (two-month intervals), seasonal, and fixed (annual mean). This combined spatial–temporal design allows us to assess how both clustering scale and temporal resolution influence the accuracy and seasonal adaptability of the correction model.

\subsection{Reanalysis-Driven Wind Power Simulation}
\label{sec2.4}

With the reanalysis datasets prepared and the spatial partition set at the desired granularity, we now perform reanalysis-driven wind power simulations to support subsequent wind speed bias correction. Our primary response variable is the capacity factor (CF), defined as the ratio of actual energy output to the theoretical maximum over the same interval:
\begin{equation}
\mathrm{CF} = \frac{E_{\mathrm{actual}}}{P_{\mathrm{rated}} \cdot \Delta t},
\end{equation}
where \(E_{\mathrm{actual}}\) is the actual energy output (in kWh or MWh), \(P_{\mathrm{rated}}\) is the turbine's rated power (in kW or MW), typically obtained from manufacturer metadata, and \(\Delta t\) is the time interval (in hours). The capacity factor is a dimensionless indicator used throughout this study to evaluate the accuracy of simulated wind power. 

The simulation workflow comprises four stages: 
(1) we extract the required variables from the reanalysis datasets, in particular wind speeds at the available reference heights; 
(2) we vertically extrapolate wind speeds to each turbine’s hub height using a logarithmic wind profile \cite{tennekes1973logarithmic} consistent with the site roughness specification;
(3) we perform spatial interpolation to each turbine or farm location and align the data to the model timestep; and 
(4) we convert the interpolated hub-height wind speeds directly to capacity factor using turbine-specific, normalized power curves described in \cite{staffell2016using}, and compile the results into capacity-factor time series. These simulated CFs are subsequently compared with observations to quantify wind-speed bias and to evaluate the effectiveness of the proposed bias corrections. Previously, this whole workflow was presented and validated using the MERRA-2 dataset in \cite{staffell2014does,staffell2015there}. Two reanalysis datasets are employed in this work: ERA5 and MERRA-2. Differences in variable availability, vertical levels, and spatial resolution (see \autoref{tab:reanalysis_comparison}) necessitate minor dataset\textendash specific handling; the overall workflow remains unchanged.

\subsubsection{Surface roughness and height extrapolation}
Surface roughness length ($z_0$) \cite{thomas1981characterization} characterises the influence of terrain on the wind profile within the atmospheric surface layer. 
Under neutral stability conditions, the mean horizontal wind speed follows the logarithmic profile:
\begin{equation}
u(z) = \frac{u_*}{\kappa} \ln\left(\frac{z}{z_0}\right),
\label{eq:loglaw_basic}
\end{equation}
where $u(z)$ is the wind speed at height $z$, $u_*$ is the friction velocity (a scaling parameter representing the shear stress at the surface), and $\kappa \approx 0.4$ is the von K\'arm\'an constant \cite{andreas2006evaluations}.  

Given wind speed measurements at two known heights $z_a$ and $z_b$, as provided by ERA5 and MERRA reanalysis products in \autoref{tab:reanalysis_comparison}, $z_0$ can be obtained analytically by eliminating $u_*$ from the log-law:
\begin{equation}
z_0 = \exp\left( \frac{u(z_b) \ln z_a - u(z_a) \ln z_b}{u(z_b) - u(z_a)} \right),
\label{eq:z0_formula}
\end{equation}
where $u(z_a)$ and $u(z_b)$ are the wind speed magnitudes derived from the eastward and northward components at the respective heights. This formulation assumes horizontally homogeneous terrain and neutral stratification, conditions under which the log-law is widely applied in wind resource assessment \cite{stull1988introduction,troen1989european}.

Once the surface roughness length $z_0$ has been estimated, the wind speed at turbine hub height $h$ is calculated as
\begin{equation}
U_{\mathrm{sim}}(h) = U(z_b) \,
\frac{\ln \left( h / z_0 \right)}{\ln \left( z_b / z_0 \right)},
\label{eq:height_extrapolation}
\end{equation}
where $U_{\mathrm{sim}}(h)$ denotes the simulated wind speed at hub height $h$, $U(z_b)$ is the reanalysis wind speed at the reference height $z_b$, $z_0$ is the surface roughness length in metres, and both $h$ and $z_b$ are expressed as metres above ground level.

\subsubsection{Spatial interpolation method}

After computing hub-height winds via \eqref{eq:height_extrapolation}, we map them from the native reanalysis grid to the exact turbine coordinates using linear interpolation in the horizontal plane. For each turbine located at longitude $\lambda_t$ and latitude $\phi_t$ with hub height $h_t$, we evaluate $U_{\mathrm{sim}}(h_t,\lambda_t,\phi_t)$ by applying separable linear interpolation over $(\lambda,\phi)$ on the gridded field $U_{\mathrm{sim}}(h,\lambda_g,\phi_g)$ and then indexing the height dimension at $h=h_t$ (i.e., no additional vertical interpolation at this stage). 

In practice, this is implemented in Python using \texttt{xarray.DataArray.interp} along longitude and latitude. Out-of-domain targets are not extrapolated; instead, we retain missing values in the dataset. Retaining missing values at out-of-domain locations ensures that no spurious wind speed estimates are introduced in areas without reanalysis coverage, thereby preventing downstream bias in capacity factor simulations. Linear interpolation offers a balance between computational efficiency and accuracy for gridded reanalysis data, avoiding oscillations that may occur with higher-order schemes when applied to sparse or irregular turbine locations. By matching ERA5 and MERRA-2 wind fields to the exact site coordinates, this step supplies location-specific hub-height wind speeds that serve as inputs to the subsequent power curve–based simulation of capacity factors.

\subsection{Bias Correction of Reanalysis Wind Speeds}
\label{sec 2.5}
Unlike purely meteorological analyses \cite{decker2012evaluation}, our bias assessment is based on the power output derived from wind farms, rather than directly on reanalysis wind speeds. This choice reflects the fact that, for energy system modelling and planning, power generation is the quantity of primary interest, while wind speed serves as an intermediate variable. Once the bias is quantified in terms of simulated output, the correction is applied back to the underlying wind speeds. This approach rests on the assumption that the dominant source of error lies in the wind speed estimates rather than in the power conversion procedure \cite{staffell2016using}. By correcting the wind speeds, we maintain consistency across all turbines and avoid unrealistic outcomes such as simulated capacity factors exceeding their physical limit of 100\%. The capacity factor bias, $B_{\mathrm{CF}}$, is defined as
\begin{equation}
B_{\mathrm{CF}} = \frac{\mathrm{Observed \ Capacity \ Factor (CF_{obs})}}{\mathrm{Simulated \ Capacity \ Factor (CF_{sim})}},
\label{eq:cf_bias}
\end{equation}
where the observed capacity factor is derived from power generation data introduced in Section \ref{sec2.2}, and the simulated capacity factor is obtained using the method proposed in Section \ref{sec2.4} driven by reanalysis wind speeds. 

\subsubsection{Wind Speed Correction Formulation}
To account for systematic biases in reanalysis wind speed data, we apply a linear correction to the raw wind speed at each site. Wind speeds are corrected using both a scalar ($\alpha$, multiplicative factor) and an offset ($\beta$, additive term):

\begin{equation}
U_{\text{corr}} = \alpha U_{\text{raw}} + \beta,
\label{eq:linear_correction}
\end{equation}

\noindent where \( U_{\text{raw}} \) is the original reanalysis wind speed (e.g., from ERA5 or MERRA-2), and \( U_{\text{corr}} \) is the corrected wind speed used for power simulation. The scalar \( \alpha \) acts as a multiplicative correction factor (e.g., under- or overestimation of wind speed magnitudes), while \( \beta \) is an additive correction term applied to the wind speed.

For each spatial cluster \(c\) and each temporal aggregation period \(\tau\) in year \(y\) as introduced in Section \ref{sec2.3}, 
we correct the raw reanalysis wind speed using a linear model with a multiplicative scalar 
\(\alpha_{y,\tau,c}\) and an additive offset \(\beta_{y,\tau,c}\).
Below, we illustrate the formulation for the case where the temporal aggregation is monthly
(\(\tau = m\)); the same approach applies to bimonthly, seasonal, or fixed groupings:
\begin{equation}
U^{\mathrm{corr}}_{j,t} \;=\; \alpha_{y,m,c}\,U^{\mathrm{raw}}_{j,t} \;+\; \beta_{y,m,c},
\label{eq:linear_correction_month}
\end{equation}
where \(U^{\mathrm{raw}}_{j,t}\) denotes the original reanalysis wind speed (e.g., ERA5 or MERRA-2) for turbine \(j\) at time step \(t\) within month \(m\), and \(U^{\mathrm{corr}}_{j,t}\) is the corrected wind speed used for subsequent power simulations.

Given a turbine-specific power curve \(C_j(\cdot)\) mapping wind speed to capacity factor (CF) in \([0,1]\), the simulated CF at \((j,t)\) is
\begin{equation}
CF_{j,t}(\alpha,\beta) \;=\; C_j\!\big(U^{\mathrm{corr}}_{j,t}(\alpha,\beta)\big).
\label{eq:pc_mapping_month}
\end{equation}

To compare simulations with observations at the monthly scale, we aggregate CFs across turbines in cluster \(c\) using capacity weights and then average over all time steps in month \(m\):
\begin{equation}
\overline{CF}^{\mathrm{sim}}_{y,m,c}(\alpha,\beta)
\;=\;
\frac{1}{\lvert \mathcal{T}_{y,m}\rvert}\,
\sum_{t\in \mathcal{T}_{y,m}}
\left(
\frac{\sum_{j\in J_c} w_j\, CF_{j,t}(\alpha,\beta)}{\sum_{j\in J_c} w_j}
\right),
\label{eq:monthly_agg}
\end{equation}
where \(\mathcal{T}_{y,m}\) is the set of time steps (e.g., hourly records) in \((y,m)\); \(J_c\) is the set of turbines assigned to cluster \(c\); \(w_j\) denotes the capacity weight (e.g., nameplate capacity in MW) for turbine \(j\), and \(\sum_{j\in J_c}w_j\) normalizes the within-cluster average.

We estimate the monthly multiplicative factor by comparing observed and simulated CFs computed from the \emph{raw} wind speeds (i.e., \(\alpha=1,\beta=0\)):
\begin{equation}
\overline{CF}^{\mathrm{sim}}_{y,m,c}
\;\equiv\;
\overline{CF}^{\mathrm{sim}}_{y,m,c}(1,0),
\qquad
\alpha_{y,m,c}
\;=\;
\frac{\overline{CF}^{\mathrm{obs}}_{y,m,c}}
     {\overline{CF}^{\mathrm{sim}}_{y,m,c}},
\label{eq:alpha_ratio_month}
\end{equation}
where \(\overline{CF}^{\mathrm{obs}}_{y,m,c}\) is the observed monthly CF for \((y,m,c)\), aligned in time with \(\overline{CF}^{\mathrm{sim}}_{y,m,c}\) and aggregated using the same capacity-weighted scheme. The ratio in \eqref{eq:alpha_ratio_month} corrects multiplicative bias: if simulations underestimate on average, \(\alpha_{y,m,c}>1\) scales speeds up; if they overestimate, \(\alpha_{y,m,c}<1\) scales them down.

With \(\alpha\) fixed at \(\alpha_{y,m,c}\), the offset \(\beta\) is determined by a one-dimensional least-squares problem:
\begin{equation}
\beta^\star_{y,m,c}
\;=\;
\arg\min_{\beta\in\mathbb{R}}
\left[
\overline{CF}^{\mathrm{sim}}_{y,m,c}\!\big(\alpha_{y,m,c},\beta\big)
-\overline{CF}^{\mathrm{obs}}_{y,m,c}
\right]^2.
\label{eq:beta_ls_month}
\end{equation}
Substituting \(\alpha_{y,m,c}\) and \(\beta^\star_{y,m,c}\) back into \eqref{eq:linear_correction_month} yields the corrected wind speeds for all \(t\in\mathcal{T}_{y,m}\) and \(j\in J_c\), thereby completing the calculation of both correction parameters.

To improve numerical stability and interpretability of the model, particularly in spatially aggregated regression settings, we reformulate 
Equation~\eqref{eq:linear_correction} into a mean-centered expression using 
the average wind speed \( \bar{U} \) over the training period. 
We define
\begin{align}
\tilde{U}_{\text{corr}} &:= U_{\text{corr}} - \bar{U},
\label{eq:def_tilde_Ucorr} \\[1ex]
\tilde{U}_{\text{raw}} &:= U_{\text{raw}} - \bar{U},
\label{eq:def_tilde_Uraw}
\end{align}
\noindent where $\tilde{U}$ indicates a mean-centered variable. Substituting these definitions into Equation~\eqref{eq:linear_correction} gives
\begin{align}
\tilde{U}_{\text{corr}} 
&= \alpha (U_{\text{raw}} - \bar{U}) + \big[\beta + (\alpha - 1)\bar{U}\big] 
= \tilde{\alpha}\,\tilde{U}_{\text{raw}} + \tilde{\beta}.
\label{eq:mean_centered}
\end{align}

\noindent Equation~\eqref{eq:mean_centered} presents the mean-centered form of the correction. In this formulation, the original correction parameters are the scalar \(\alpha\) (multiplicative factor) and the offset \(\beta\) (additive term). Their mean-centered forms are denoted as the stretch factor \(\tilde{\alpha}\) and the shift factor \(\tilde{\beta}\). Here, \(\alpha (U_{\text{raw}} - \bar{U})\) scales deviations from the local mean wind speed, while the composite term [\(\beta + (\alpha - 1)\bar{U}\)] adjusts the site-specific mean. Such a formulation supports stable parameter estimation across diverse wind farm sites. It also provides a flexible basis for exploring the relationships between wind speed bias and geographic features. By separating the mean and deviation components, it improves comparability across regions and interpretability of regional bias patterns.

\subsubsection{Performance Evaluation Metrics}
\label{sec 2.5.2}
We evaluate model performance in the test year \(y=2019\) at four temporal granularities: monthly, bimonthly, seasonal, and fixed. 
Let \(\tau\in\{\text{month},\ \text{bimonth},\ \text{season},\ \text{fixed}\}\). 
For a given \(\tau\), let \(\mathcal{I}_{\tau}\) denote the set of all cluster–time-slice pairs \((c,k)\) in 2019 at granularity \(\tau\), where \(k=1,\ldots,K_\tau\) with \(K_{\text{month}}=12\), \(K_{\text{bimonth}}=6\), \(K_{\text{season}}=4\), and \(K_{\text{fixed}}=1\). 
For each \(i\equiv(c,k)\in\mathcal{I}_{\tau}\), we obtain the capacity factors $CF^{\mathrm{sim}}_{i}$ and $CF^{\mathrm{obs}}_{i}$, and the error is defined as
\[
e_i \;\equiv\; CF^{\mathrm{sim}}_{i} - CF^{\mathrm{obs}}_{i},
\]
where \(CF^{\mathrm{sim}}_{i}\) and \(CF^{\mathrm{obs}}_{i}\) are the capacity-weighted CFs for cluster \(c\) in the \(k\)-th time slice of 2019, aggregated using the same scheme as in the modelling section. All CFs are dimensionless in \([0,1]\).

With \(N=\lvert\mathcal{I}_{\tau}\rvert\), the metrics reported for each \(\tau\) are
\begin{align}
\mathrm{RMSE}(CF^{\mathrm{sim}},CF^{\mathrm{obs}})
&= \sqrt{\frac{1}{N}\sum_{i=1}^{N} e_i^2}, \label{eq:rmse_2019} \\
\mathrm{MAE}(CF^{\mathrm{sim}},CF^{\mathrm{obs}})
&= \frac{1}{N}\sum_{i=1}^{N} \lvert e_i\rvert, \label{eq:mae_2019} \\
\mathrm{MBE}(CF^{\mathrm{sim}},CF^{\mathrm{obs}})
&= \frac{1}{N}\sum_{i=1}^{N} e_i. \label{eq:mbe_2019}
\end{align}
The root-mean-square error (RMSE) quantifies the average magnitude of the errors using a quadratic scoring rule, which gives greater weight to larger deviations and is therefore more sensitive to outliers. The mean absolute error (MAE) also measures the average magnitude of the errors, but uses a linear scoring rule that treats all deviations equally, making it less sensitive to extreme values. The mean bias error (MBE) is the signed mean error, where a positive value indicates systematic overestimation and a negative value indicates systematic underestimation. All three metrics attain their optimal value at zero, with smaller magnitudes indicating higher accuracy; for MBE, values close to zero indicate minimal systematic bias.

\subsection{Geospatial Correlation Analysis}
\label{Sec 2.6}
We assess the pairwise association between the site-level wind-speed bias correction factor and some geospatial features using Pearson correlation $r$ \cite{schober2018correlation} and Spearman correlation $\rho$ \cite{deWinter2016}.

\paragraph{Pearson correlation}
Given paired observations $\{(x_i, y_i)\}_{i=1}^n$ (a geospatial feature $x_i$ and the correction factor $y_i$ at site $i$), Pearson’s $r$ measures the strength of a \emph{linear} association between $x$ and $y$:
\begin{equation}
r \;=\;
\frac{\sum_{i=1}^{n} (x_i-\bar{x})(y_i-\bar{y})}
{\sqrt{\sum_{i=1}^{n} (x_i-\bar{x})^2}\;
 \sqrt{\sum_{i=1}^{n} (y_i-\bar{y})^2}},
\quad
\bar{x}=\frac{1}{n}\sum_{i=1}^{n} x_i,\;
\bar{y}=\frac{1}{n}\sum_{i=1}^{n} y_i.
\label{eq:pearson}
\end{equation}
The coefficient $r$ takes values in $[-1,1]$, with $\pm1$ denoting perfect positive or negative linear association and $0$ no association. Pearson’s $r$ is sensitive to outliers because it relies directly on deviations from the mean.

\paragraph{Spearman correlation}
Spearman’s $\rho$ is a \emph{rank-based} analogue of Pearson’s $r$. It applies \eqref{eq:pearson} to the ranks of $x_i$ and $y_i$:
\begin{equation}
\rho \;=\; \mathrm{corr}\!\big(R(x),\,R(y)\big),
\label{eq:spearman}
\end{equation}
where $R(\cdot)$ denotes mid-rank assignment (average ranks for ties). Rather than measuring strictly linear association, Spearman’s $\rho$ measures monotonic association and takes values in $[-1,1]$. Because it is based on ranks, Spearman’s $\rho$ is more robust to outliers.

\paragraph{Significance testing}
To assess the statistical significance of the correlations, we report \(p\)-values. 
For Pearson’s \(r\), the null hypothesis \(H_{0}\!:\!r=0\) is tested using a \(t\)-statistic:
\begin{equation}
t = r \sqrt{\frac{n-2}{1-r^2}}, \qquad \mathrm{df} = n-2,
\end{equation}
which follows a \(t\)-distribution with \(n-2\) degrees of freedom. 
For Spearman’s \(\rho\), \(p\)-values are obtained from large-sample or permutation approximations.

\paragraph{Confidence intervals}
To quantify the uncertainty of the estimated correlations, we additionally report 95\% confidence intervals (CIs) for Pearson’s $r$ \cite{artusi2002bravais,de2016comparing,schober2018correlation}, 
\begin{equation}
z = \tfrac{1}{2}\ln\!\left(\frac{1+r}{1-r}\right), 
\qquad \mathrm{SE}_z = \frac{1}{\sqrt{n-3}},
\end{equation}
For a two-sided $(1-\alpha)$ interval, the bounds for $z$ are given by
\begin{equation}
z \pm z_{\alpha/2}\,\mathrm{SE}_z,
\label{eq:z_interval}
\end{equation}
with two-sided bounds \(z\pm z_{\alpha/2}\mathrm{SE}_z\) (where \(z_{\alpha/2}\) is the standard normal quantile; 1.96 for \(\alpha=0.05\)). 
The limits are then back-transformed (via \(\tanh\)) to obtain CIs for \(r\). 
CIs are computed only for Pearson’s \(r\); for Spearman’s \(\rho\) we report the coefficient and its \(p\)-value.

\paragraph{Multiple regression analysis}
To complement the pairwise correlation analysis, we applied a multiple linear regression model \cite{cohen2013applied} with the correction factor $y_i$ as the dependent variable and a set of geospatial predictors $\mathbf{x}_i = (x_{1,i}, x_{2,i}, \ldots, x_{p,i})$ as independent variables:
\begin{equation}
y_i = \beta_0 + \sum_{j=1}^{p} \beta_j x_{j,i} + \varepsilon_i,
\label{eq:mlr}
\end{equation}
where $\varepsilon_i$ denotes the residual error term associated with site $i$, and $p$ is the number of predictors. 
The regression coefficients $\beta_j$ were estimated by ordinary least squares (OLS), and their statistical significance was evaluated using \(t\)-statistics and corresponding two-tailed \(p\)-values.

\paragraph{Geospatial features}
We consider four features:
\begin{itemize}
  \item \textbf{Elevation}: height above sea level (m).
  \item \textbf{Hilliness}: difference between the highest and lowest elevation within a 10\,km radius (m).
  \item \textbf{Local Turbine Count}: number of wind turbines within a 7.5\,km radius (count).
  \item \textbf{Distance to sea}: geodesic distance from the site to the nearest sea coastline (km).
\end{itemize}
The datasets for elevation and hilliness were obtained from two complementary sources. Elevation at turbine coordinates was retrieved through the Google Maps Elevation API \cite{google_elevation_api,wang2017google,zhu2018driving}, which provides efficient real-time access to elevation data but is quota-limited and not open source. Hilliness was calculated from all elevation grid cells contained in a 10 km-radius buffer around each turbine. To assess the reliability of these estimates, values were cross-checked against the Copernicus GLO-30 digital elevation model accessed via Google Earth Engine (GEE) \cite{google_earthengine_copernicus_dem,cop_dem_product_handbook,gorelick2017google,mutanga2019google}, indicating correspondence between the two datasets. Copernicus GLO-30 is a 30\,m global digital elevation model (DEM) developed under the Copernicus programme \cite{thepaut2018copernicus} and is openly accessible for scientific use. In this study, we additionally employed the Copernicus GLO-30 DEM to produce a national-scale elevation map of the UK, which is used in subsequent analyses. The UK boundary shapefile was obtained from the Natural Earth dataset 
(\texttt{ne\_10m\_admin\_0\_countries}, 1:10m cultural vectors; \cite{naturalearth}). Distance to the sea was calculated as the geodesic distance from each site to the nearest coastline, approximated by the distance to the closest boundary coordinate of the UK shoreline. Local turbine count was defined as the number of wind turbines within a 7.5\,km radius of each site. This measure reflects the degree of local concentration, which may influence wake interactions and in turn bias reanalysis wind-speed estimates. Coherent turbine wake structures have been observed at least 30 rotor diameters downstream in onshore settings \cite{uchida2022effects}, while wake effects offshore are known to persist to even greater distances, and are projected to grow with increasing build-out \citep{warder2025future}. Such findings guided our choice of a radius consistent with our dataset, encompassing both onshore and offshore turbines and therefore accommodating differences in rotor size and wake extent. Turbine geographic coordinates were primarily obtained from OpenStreetMap \cite{openstreetmap,mooney2017review,haklay2008openstreetmap}, which provides comprehensive geospatial coverage throughout the UK, and were supplemented with the official UK government dataset Renewable Energy Planning Database (REPD) \cite{repd_data_gov_uk}. The metadata contained in the dataset described in Section~\ref{sec2.2} also provides some supplementary details.

\section{Results and discussion}
\label{sec:3}

\subsection{Performance of the Bias Correction Model}
This section presents the performance of the bias-correction model. Wind farms are first grouped into spatially coherent clusters, and simulations are performed at multiple temporal aggregations as described in Section \ref{sec2.3}. Model improvements are quantified using the metrics defined in Section \ref{sec 2.5}: the root-mean-square error (RMSE), mean absolute error (MAE), and mean bias error (MBE), each calculated as the discrepancy between simulated and observed capacity factors in the test year. We begin by applying the full pipeline to ERA5, which is the primary reanalysis dataset considered in this study and represents the main application of our correction model. To examine cross-dataset applicability, we then apply exactly the same clustering, simulation, and evaluation steps to MERRA-2. Finally, we compare the two sets of results to assess consistency in error reduction across datasets and to identify any systematic differences attributable to the underlying reanalyses. Together, these analyses quantify reductions in random error (RMSE/MAE) and systematic bias (MBE) across clustering schemes and temporal resolutions, providing evidence for the generalisability of the proposed correction model.

\subsubsection{Evaluation of ERA5 Correction Results}
\begin{figure}[htbp]
    \centering
    \includegraphics[width=1\textwidth]{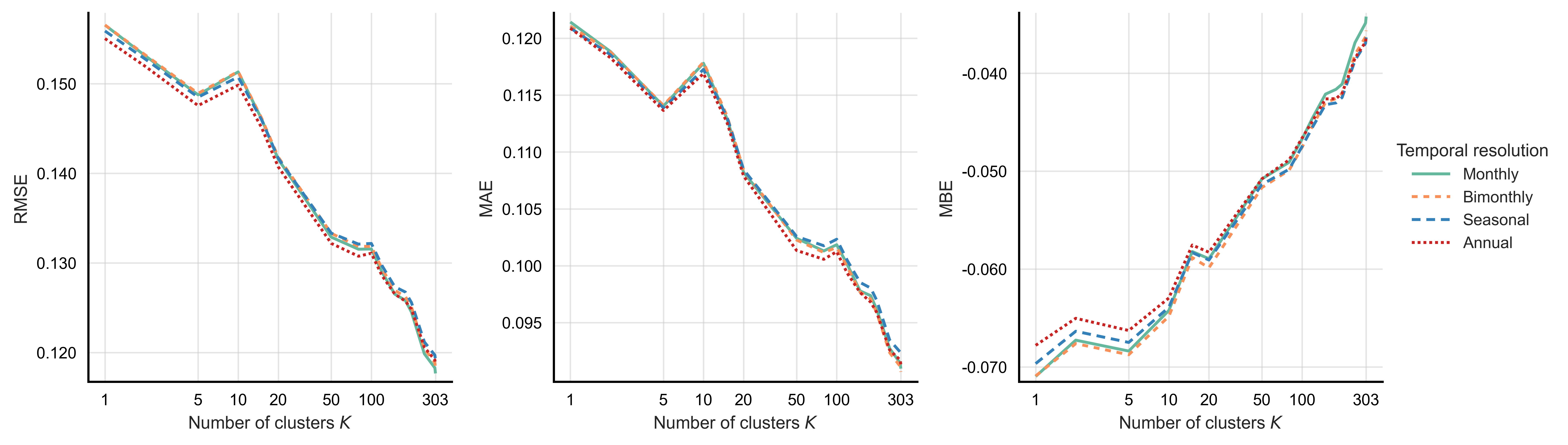}
    \caption{Impact of clustering and temporal resolution on three error metrics (RMSE, MAE, and MBE) for ERA5.}
    \label{fig5}
\end{figure}

As shown in \autoref{fig5}, increasing the number of spatial clusters ($K$) leads to a clear improvement in all error metrics for the ERA5 dataset. Both the RMSE and MAE exhibit a clear decreasing trend with larger $K$, despite minor fluctuations at intermediate values, indicating that finer spatial granularity in bias correction reduces random errors. With a single cluster ($K$ = 1), RMSE and MAE are relatively high, reflecting larger discrepancies. As $K$ grows, these discrepancies diminish: the RMSE and MAE curves drop substantially from their initial values. Similarly, MBE also improves (moves closer to zero) as the number of clusters increases across all temporal resolutions. Notably, MBE remains negative in all scenarios (a negative MBE indicates that the model tends to systematically underestimate the observed capacity factor), but the magnitude of this bias shrinks with more clusters. This means that a higher $K$ improves precision and mitigates systematic bias: the initially pronounced underestimation bias is progressively corrected. This trend highlights that spatial clustering of the correction factors effectively captures regional variations in error, reducing both random errors and consistent underestimation in the model output. 

The choice of temporal resolution exerts only a modest influence on the error metrics based on simulated–observed CF differences. In \autoref{fig5}, for any given $K$, the curves for monthly (green, solid line), bimonthly (orange, dashed line), seasonal (blue, dashed line), and annual (red, dotted line) corrections lie close to one another, and the between-resolution spread is small compared with the reductions achieved by increasing $K$. At small to intermediate cluster counts (approximately $K \leq 100$), the annual correction frequently attains RMSE and MAE values that are as low as or slightly lower than those of the higher-frequency schemes, indicating that a single time-invariant factor can be adequate when spatial granularity is coarse. Once the cluster count is large ($K > 100$), monthly and bimonthly updates begin to marginally outperform the annual scheme, yielding the lowest RMSE and MAE. MBE shows a similar pattern across different resolutions: at high $K$, monthly and bimonthly corrections bring it slightly nearer to zero than annual or seasonal updates.

By incorporating spatially granular corrections and frequent temporal updates, the corrected ERA5 outputs show lower RMSE and MAE compared to uncorrected simulations, indicating a much closer agreement with observed capacity factor values. Overall, the most accurate results are obtained when using $K = 303$ clusters in combination with monthly bias updates, which reflects the highest agreement with the observed data. The reductions in RMSE and MAE, together with the mitigation of systematic bias (MBE shifting toward zero), indicate the effectiveness of the correction methodology. Although a small residual underestimation remains, the model achieves higher accuracy by lowering random errors and mitigating persistent bias, leading to more reliable ERA5-based capacity factor estimates.

\subsubsection{Evaluation of MERRA-2 Correction Results}

\begin{figure}[htbp]
    \centering
    \includegraphics[width=1\textwidth]{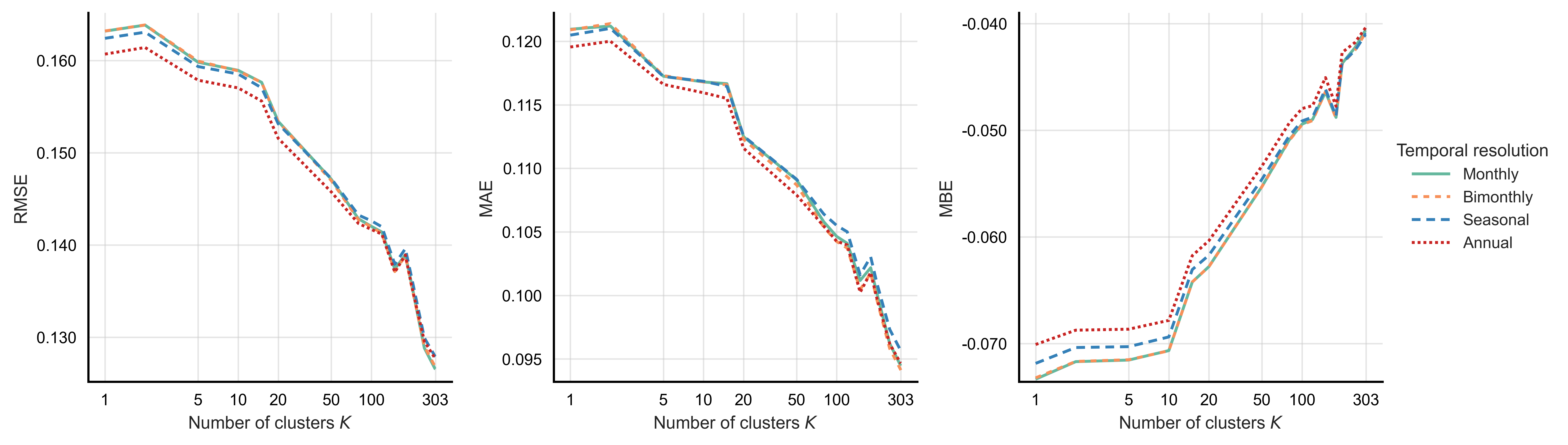}
    \caption{Impact of clustering and temporal resolution on error metrics (RMSE, MAE, and MBE) for MERRA-2.}
    \label{fig6}
\end{figure}

For MERRA-2, the correction exhibits the same qualitative behaviour as for ERA5. In \autoref{fig6}, increasing the number of clusters $K$ leads to declines in both RMSE and MAE, while MBE remains negative yet moves toward zero, indicating an attenuated (but not eliminated) underestimation. Therefore, the correction results for MERRA-2 display a pattern very similar to those for ERA5 with respect to the influence of cluster number, suggesting that the overall response of the metrics is consistent across the two reanalysis datasets. Regarding temporal resolution, MERRA-2 appears to be more sensitive than ERA5, as the separation between the four curves (monthly, bimonthly, seasonal, and annual) is more pronounced across all error metrics. MERRA-2 often achieves lower overall errors under the annual correction scheme, particularly at smaller cluster numbers. However, for RMSE and MAE, once the number of clusters becomes sufficiently large, particularly beyond $K = 150$, the monthly and bimonthly corrections again yield the lower errors. This behaviour is broadly consistent with ERA5, suggesting that under finely resolved spatial clustering, higher-frequency temporal updates confer additional benefits in both reanalysis datasets. By contrast, in terms of systematic error (MBE), finer temporal resolutions provide little advantage over the annual scheme. These results indicate that the proposed correction framework is transferable across reanalysis products, effectively improving both random error and systematic bias.

\subsubsection{Quantitative Comparison of ERA5 and MERRA-2 Corrections Results}

\begin{figure}[htbp]
    \centering

    \begin{subfigure}{0.24\textwidth}
        \centering
        \includegraphics[width=\linewidth]{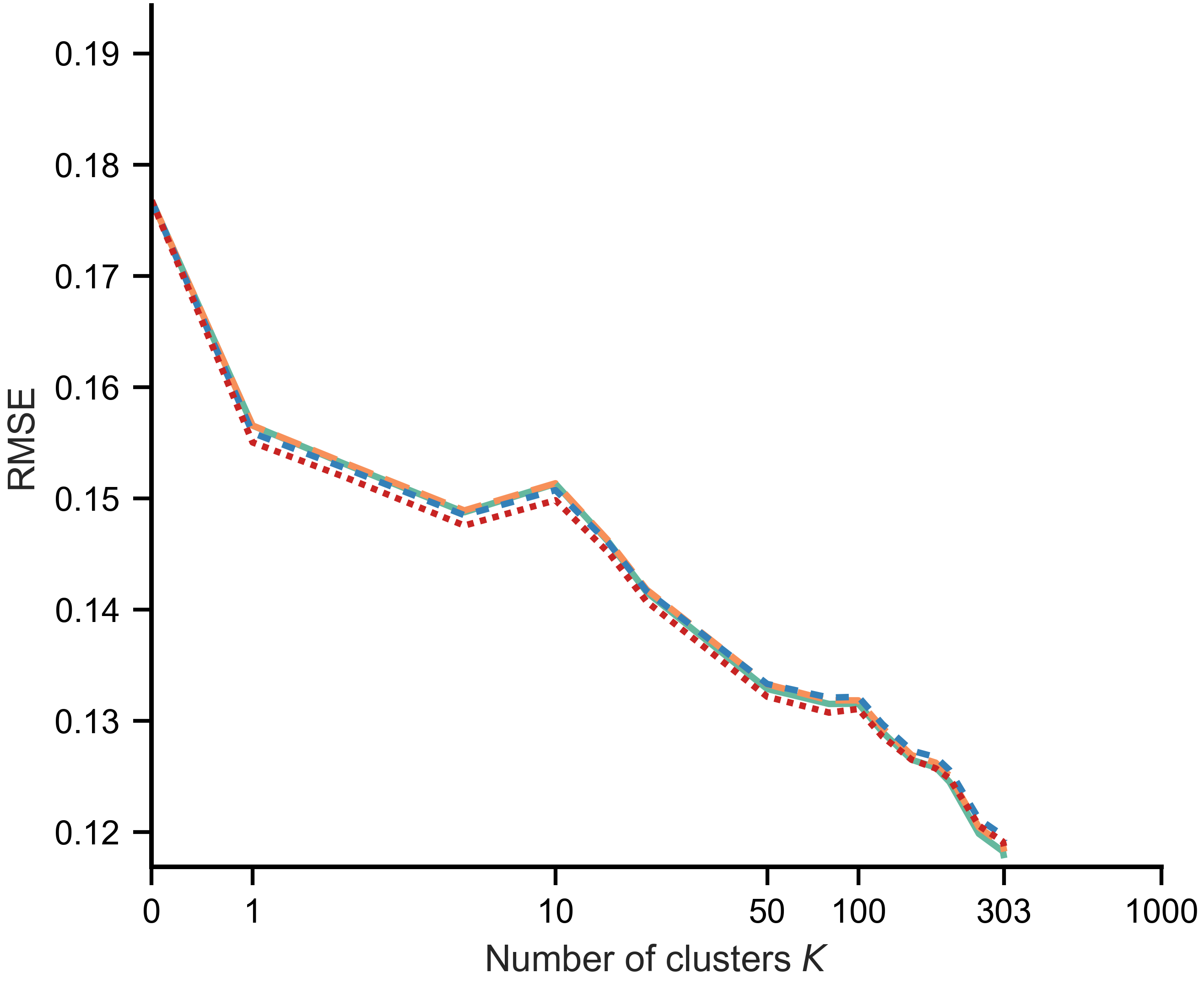}
        \caption{ERA5 RMSE}
        \label{fig7a}
    \end{subfigure}\hfill
    \begin{subfigure}{0.24\textwidth}
        \centering
        \includegraphics[width=\linewidth]{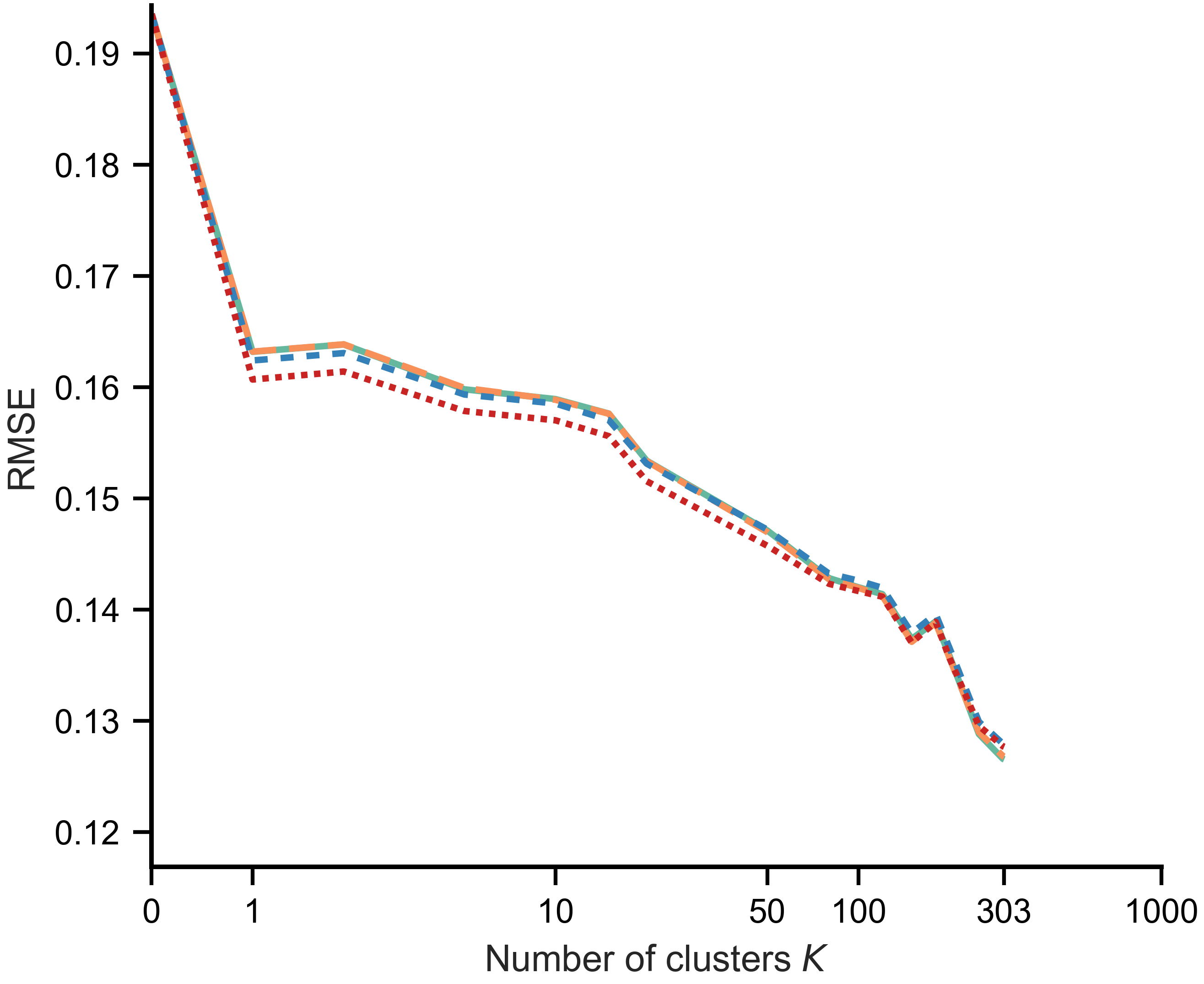}
        \caption{MERRA-2 RMSE}
        \label{fig7b}
    \end{subfigure}\hfill
    \begin{subfigure}{0.24\textwidth}
        \centering
        \includegraphics[width=\linewidth]{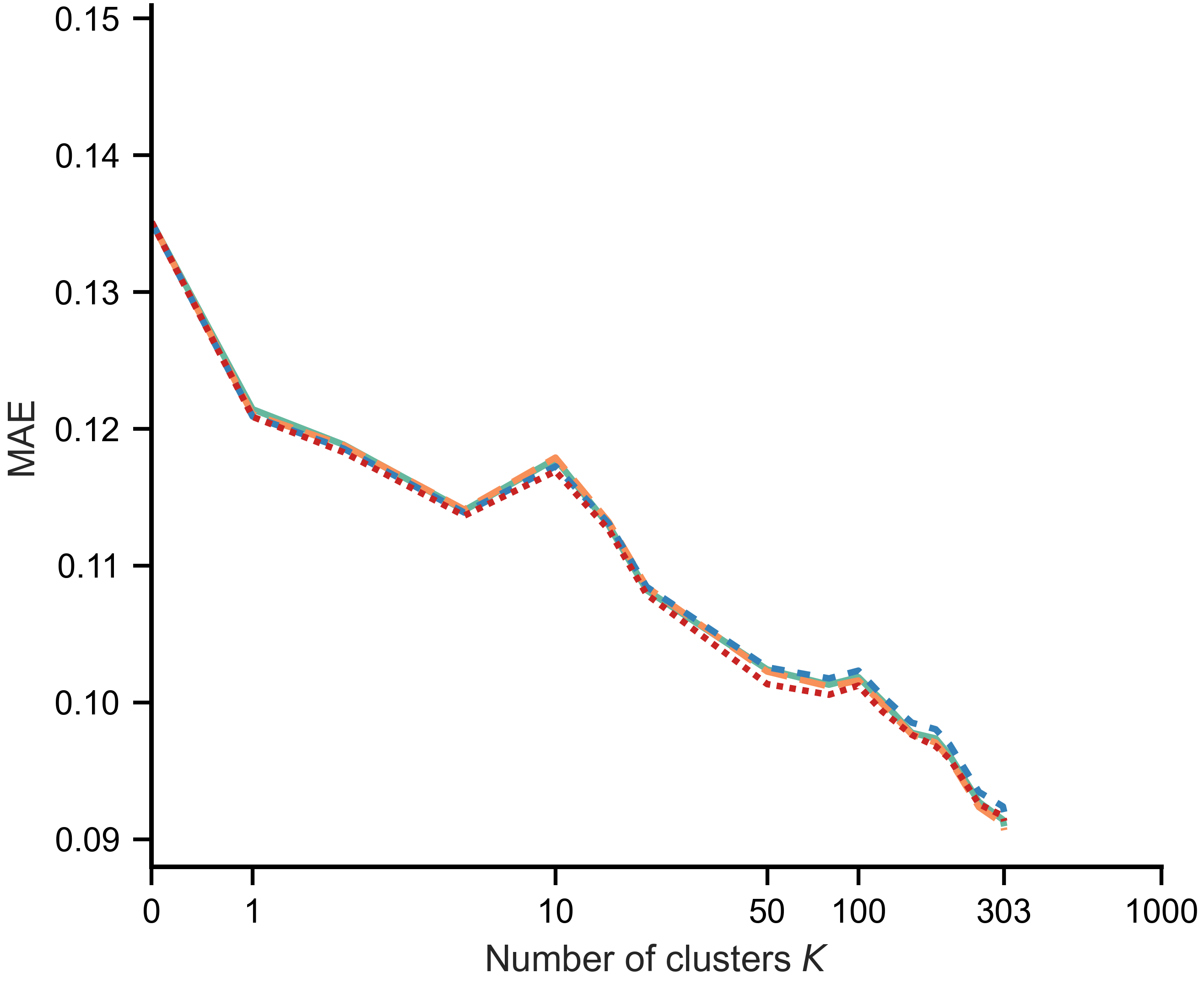}
        \caption{ERA5 MAE}
        \label{fig7c}
    \end{subfigure}\hfill
    \begin{subfigure}{0.24\textwidth}
        \centering
        \includegraphics[width=\linewidth]{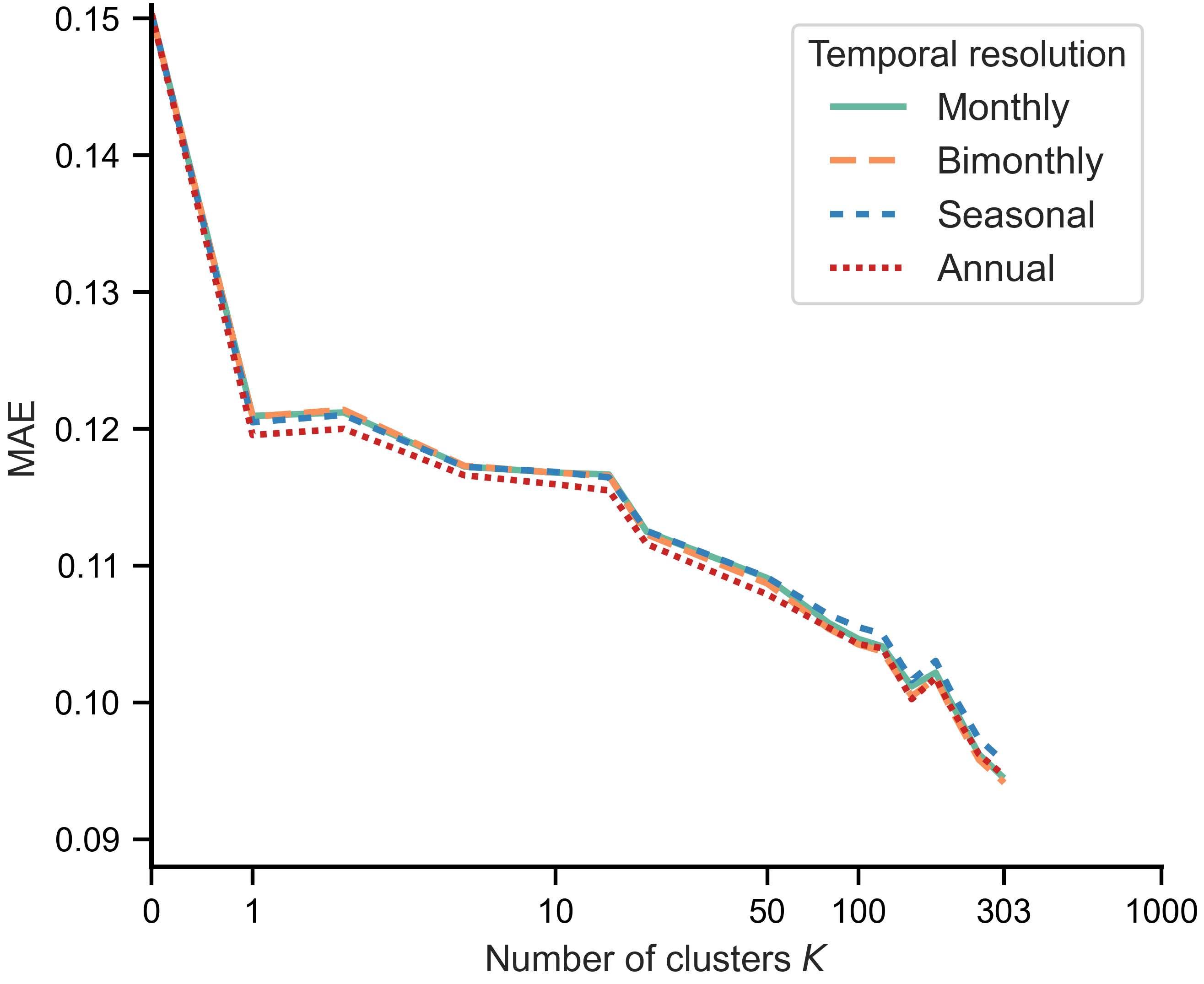}
        \caption{MERRA-2 MAE}
        \label{fig7d}
    \end{subfigure}

    \caption{Comparison of bias correction results between ERA5 and MERRA-2 across different clustering and temporal resolutions (including uncorrected value at $K = 0$). Panels (a) and (b) show RMSE results for ERA5 and MERRA-2, respectively, while panels (c) and (d) present the corresponding MAE results.}
    \label{fig7}
\end{figure}

For a more direct comparison of the two reanalysis datasets and to better illustrate the improvements achieved by the correction, we introduce an uncorrected baseline $K$ = 0, representing simulations driven by the raw reanalysis wind speed data without any bias correction. In \autoref{fig7}, we plot the evolution of RMSE and MAE for ERA5 and MERRA-2 across cluster granularities including the uncorrected baseline. For ERA5 (Figure~\ref{fig7}\subref{fig7a},~\subref{fig7c}), both metrics start from relatively high values and show an overall downward trend as the number of clusters increases. The sharp drop from $K$ = 0 to $K$ = 1 corresponds to applying a single national-scale correction factor, following the approach in \cite{staffell2016using}. Applying one uniform correction factor substantially reduces both RMSE and MAE, as most of the reanalysis error is a country-wide systematic bias. This large-scale component is effectively removed by the uniform correction, explaining the step between $K=0$ and $K=1$. Subsequent increases in $K$ then refine the correction by addressing more localized, region-specific deviations, resulting in further but more gradual decreases in error. For MERRA-2 (Figure~\ref{fig7}\subref{fig7b},~\subref{fig7d}), the curves exhibit a similar downward trend, but with several notable differences. First, the raw errors at $K=0$ are larger, indicating a stronger initial bias in the uncorrected MERRA-2 data. Second, the error reduction from $K=0$ to $K=1$ is even more pronounced than in ERA5, showing a greater benefit from removing the nationwide systematic bias. Third, despite these improvements, MERRA-2 errors remain higher than those of ERA5, both before correction, throughout the clustering process, and even at the optimal correction level ($K=303$). This suggests that although the correction framework substantially improves both datasets, ERA5 retains a clear advantage in absolute accuracy across all levels of clustering. This observation is consistent with the higher spatial resolution of ERA5 compared with MERRA-2, yielding more reliable inputs for wind energy modelling. Consistent findings  have been reported in previous studies, confirming that ERA5 generally outperforms MERRA-2 in reproducing wind resources and capacity factor dynamics \cite{gruber2022towards,taszarek2021comparison,kara2024evaluation}.

To complement the graphical trends in \autoref{fig7}, the corresponding values for ERA5 and MERRA-2 are summarised in \autoref{tab:era5_merra2_comparison}. At the uncorrected baseline ($K$ = 0), ERA5 simulations yielded an RMSE of 0.177 and an MAE of 0.135, while MERRA-2 produced notably higher errors (0.194 and 0.150, respectively). After applying the maximum clustering granularity ($K$ = 303), the errors were markedly reduced: ERA5 decreased to an RMSE of 0.119 and an MAE of 0.091, corresponding to relative reductions of about 33\% for both metrics. In comparison, MERRA-2 errors declined to 0.128 and 0.096, representing reductions of 34\% and 36\%, respectively. Across all clustering levels, ERA5 consistently achieves lower errors than MERRA-2, with RMSE and MAE being 8.8\% and 10.0\% lower at the uncorrected baseline, and 7.0\% and 5.2\% lower at the maximum clustering. Notably, ERA5’s advantage over MERRA-2 is larger for RMSE than for MAE.

\begin{table}[htbp]
\centering
\renewcommand{\arraystretch}{1.15} 
\setlength{\tabcolsep}{13.5pt} 
\caption{Quantitative comparison of correction performance for ERA5 and MERRA-2 at specific cluster counts $K$. 
Error metrics (RMSE and MAE) are computed between simulated and observed capacity factors.}
\label{tab:era5_merra2_comparison}
\begin{tabular}{ccccc}
\toprule
\multirow{2}{*}{\textbf{Clusters $K$}} & \multicolumn{2}{c}{\textbf{ERA5}} & \multicolumn{2}{c}{\textbf{MERRA-2}} \\
\cmidrule(lr){2-3} \cmidrule(lr){4-5}
 & {\textbf{RMSE}} & {\textbf{MAE}} & {\textbf{RMSE}} & {\textbf{MAE}} \\
\midrule
0 (Uncorrected) & 0.177 & 0.135 & 0.194 & 0.150 \\
20              & 0.141 & 0.107 & 0.152 & 0.112 \\
80              & 0.131 & 0.101 & 0.142 & 0.105 \\
150             & 0.127 & 0.097 & 0.138 & 0.102 \\
200             & 0.125 & 0.095 & 0.136 & 0.100 \\
303             & 0.119 & 0.091 & 0.128 & 0.096 \\
\bottomrule
\end{tabular}
\end{table}

These findings provide several key insights. ERA5 consistently outperforms MERRA-2 in predictive accuracy, with lower baseline errors before bias correction is applied. This highlights the advantage of ERA5 for wind-energy modelling. Nevertheless, the clustering-based bias-correction framework effectively reduces systematic errors in both datasets. MERRA-2, due to its larger initial bias, shows larger relative percentage improvements after correction; however, its absolute errors remain higher than those of ERA5 across all clustering levels. Differences between ERA5 and MERRA-2 are larger for RMSE than for MAE, implying that larger deviations may be better controlled in the ERA5-driven simulations, whereas average deviations are more similar. From a methodological perspective, these observations confirm the robustness of the correction framework, which improves higher-quality datasets such as ERA5, alongside lower-fidelity products like MERRA-2. This demonstrates a certain degree of cross-dataset transferability, suggesting that the approach can be applied more broadly across reanalysis products. The present analysis does not disentangle performance on existing versus newly commissioned farms, so possible overfitting to unseen sites was not explicitly tested; future work may address this by evaluating the two groups separately. Therefore, while ERA5 should be considered the preferred choice for applications requiring the highest accuracy, a sufficiently corrected MERRA-2 remains a viable alternative in cases where ERA5 is unavailable or for cross-validation between datasets.

\subsection{Spatial Distribution of Bias Correction Factors}
\label{sec 3.2}
We have previously introduced our wind speed correction formulation and corresponding bias correction factors: the stretch factor $\tilde{\alpha}$ and shift factor $\tilde{\beta}$ in Section \ref{sec 2.5}. To provide a clearer illustration of the spatial distribution of errors across the United Kingdom, particularly for ERA5 given its relatively high resolution and frequent use in wind energy modelling, we map the bias correction factors. Specifically, the factors are represented by colour shading according to their magnitudes, based on the clustering scheme with $K=303$ and aggregated at the annual timescale. In this section, we discuss the spatial distribution of the correction factors as revealed by these maps, and compare them against the elevation map of the UK in order to explore potential relationships between the two. These observations motivate us to extend the analysis by introducing additional geographical variables for further discussion.

\begin{figure}[htbp]
    \centering
    \includegraphics[width=1\textwidth]{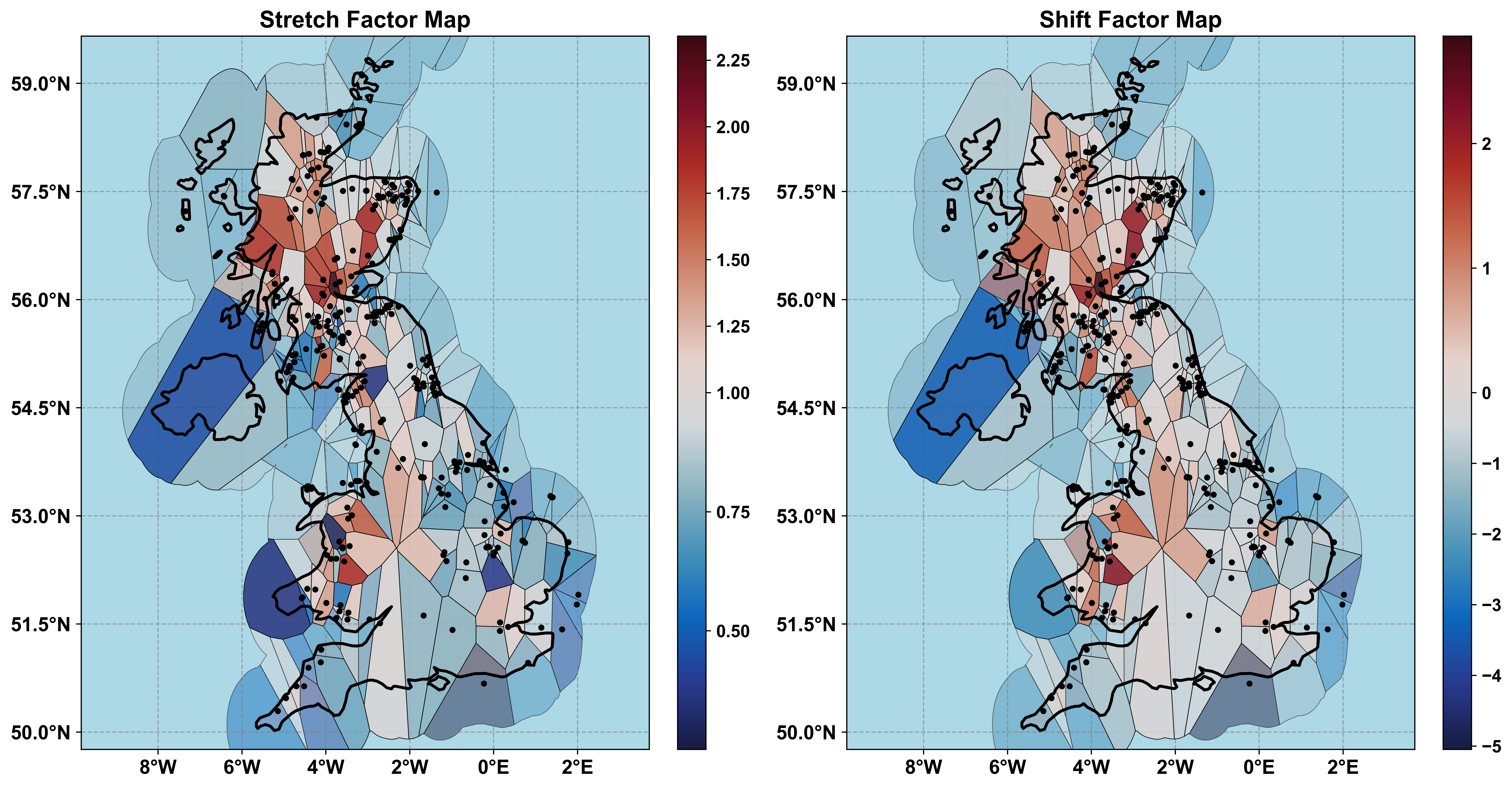}
    \caption{Bias correction factor distributions ($K=303$, annual) across UK wind farm zones. (Left) Stretch factors ($\tilde{\alpha}$), and (right) shift factors ($\tilde{\beta}$). Red indicates higher correction values, and blue indicates lower values. Because the dataset spans onshore and offshore sites, the UK coastline is overlaid for context; polygons may cross the coastline.}
    \label{fig8}
\end{figure}

The spatial distribution of correction factors in \autoref{fig8} reveals distinct wind speed bias patterns in the ERA5 dataset across different terrain types in the UK. The map partitions the United Kingdom into a set of polygons, which define the clusters formed from the constituent farms. Individual wind farms are plotted as black points within their respective cluster polygons. Polygon colours encode the magnitude of the correction factor, with larger values shown in warmer (red) hues and smaller values in cooler (blue) hues. This colour mapping enables visual comparison of relative correction magnitudes across clusters. Cluster polygons, including both onshore and offshore sites, are shown with the UK coastline overlaid for context. As formulated in \autoref{eq:mean_centered}, the two correction factors $\tilde{\alpha}$ and $\tilde{\beta}$ act in a multiplicative and an additive manner, respectively. Owing to this formulation, the two factors are inherently correlated. \autoref{fig8} illustrates that the two distributions exhibit systematic behaviour: stretch factors $\tilde{\alpha}$ exceeding unity amplify the variability of wind speeds, whereas values below unity attenuate it. In contrast, the shift factor $\tilde{\beta}$ operates additively, effecting a uniform shift in the mean level of the distribution. This implies that $\tilde{\alpha} > 1$ or $\tilde{\beta} > 0$ indicates that the reanalysis data have underestimated the wind speed at that location; conversely, $\tilde{\alpha} < 1$ or $\tilde{\beta} < 0$ indicates that the reanalysis data have overestimated the wind speed.

By examining the spatial distribution, several consistent regional patterns emerge: (1) In the Scottish Highlands, the Northwestern Islands, and Wales, ERA5 systematically underestimates wind speeds ($\tilde{\alpha} > 1$, $\tilde{\beta} > 0$, shown in red), necessitating an upward correction. (2) In Southeastern and Eastern coastal England, as well as in Western coastal England, ERA5 generally overestimates wind speeds ($\tilde{\alpha} < 1$, $\tilde{\beta} < 0$, shown in blue), requiring a downward correction. (3) Across the Central England plains and offshore regions, the stretch factor remains close to 1, indicating that ERA5 wind speed simulations are relatively accurate in these areas. These regional patterns across the UK suggest that systematic biases in ERA5 are not randomly distributed. To further investigate the drivers of these spatial patterns, we next compare the correction factors with one of the most representative topographic variables — elevation.

\begin{figure}[htbp]
    \centering
    \includegraphics[width=1\textwidth]{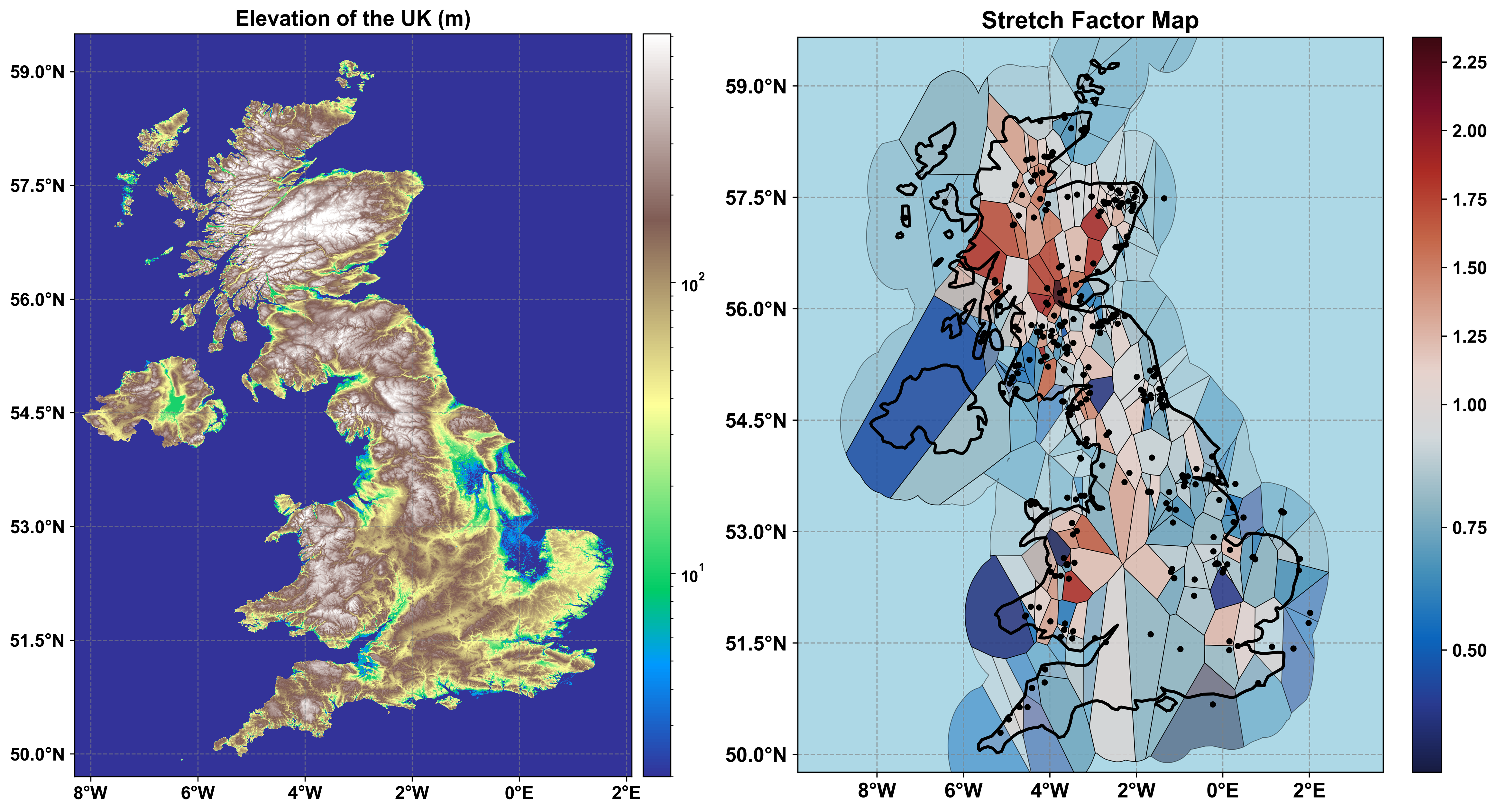}
    \caption{Comparison between UK topography and stretch-correction factor: elevation map of the UK (left, based on the Copernicus GLO-30 DEM, Google Earth Engine \cite{google_earthengine_copernicus_dem,cop_dem_product_handbook}), and spatial distribution of the stretch-correction factor $\tilde{\alpha}$ (right, based on $K=303$ clusters and annual correction).}
    \label{fig9}
\end{figure}

The elevation map of the UK with colours indicating altitude ranges is shown in the left panel of \autoref{fig9}, while the right panel shows the spatial distribution of the stretch factor $\tilde{\alpha}$. The elevation data used for plotting the map was obtained from the Copernicus Global 30\,m DEM  \cite{google_earthengine_copernicus_dem,cop_dem_product_handbook}. Given the correlation between the two correction factors discussed above, we restrict our presentation to the stretch factor $\tilde{\alpha}$, which is used for the subsequent analysis. From inspection of the maps, high-elevation regions appear concentrated in Scotland, northern England, and Wales, and they closely align with the darker red areas on the correction factor map. In certain parts of these regions, the stretch factor $\tilde{\alpha}$ and shift factor  $\tilde{\beta}$ significantly exceed 1 and 0 respectively, suggesting a systematic underestimation of wind speeds by ERA5 in elevated, mountainous, or hilly terrain. This finding is consistent with previous studies \cite{rose2016quantifying,jourdier2020evaluation,gualtieri2021reliability,potisomporn2023era5}, which have shown that ERA5 systematically underestimates wind speeds in high-elevation or topographically complex mountainous regions. Such underestimations are generally attributed to the coarse horizontal resolution of ERA5, which smooths orographic features and therefore cannot resolve local speed-up effects associated with ridges, slopes, and valleys \cite{gualtieri2021reliability}. Moreover, the vertical variation of wind speed in mountainous regions is often more pronounced than that assumed by ERA5’s boundary-layer parameterisations, leading to underestimated wind speeds at turbine hub heights. Non-linear vertical gradients \cite{potisomporn2023era5}, together with local meteorological systems such as valley winds, slope flows, and channeling effects, also remain unresolved at ERA5’s spatial scale. Collectively, these factors explain the systematic underestimation of wind speeds by ERA5 in high-elevation and topographically complex terrain.

While ERA5 systematically underestimates wind speeds in mountainous regions, the opposite tendency is observed in certain low-lying coastal areas of England. In southeastern, eastern and western coastal regions, $\tilde{\alpha}$ values are smaller than 1, in some locations even below 0.5, suggesting a pronounced overestimation of wind speeds by ERA5 in these flat terrains (requiring a downward adjustment). Comparable results were reported for Denmark \cite{benmoufok2024improving}, where wind speeds were found to be overestimated in areas located near the coastline. This overestimation is likely driven by several factors. First, the coarse horizontal resolution of ERA5 mixes land and offshore grid cells, introducing a spurious influence from stronger offshore winds. Second, its simplified surface roughness parameterisation underestimates the frictional effects of vegetation, forests, and built environments. Third, grid averaging across ~31 km cells neglects local sheltering and site-specific conditions, leading to systematically higher wind speeds in the reanalysis compared with observations.

In addition to these regional contrasts, the majority of inland plain regions in England and offshore wind farms display relatively uniform correction factors. Offshore sites in particular show correction factors largely within the range of 0.8 to 1.2, indicating that ERA5 provides relatively reliable estimates of wind speeds over flat inland areas and marine environments. This observation agrees with evidence from previous studies, which have demonstrated that errors in offshore and flat regions are generally smaller, especially in contrast to those found in mountainous and topographically complex areas \cite{gualtieri2022analysing}. This is expected, as offshore environments are characterised by relatively low and spatially homogeneous aerodynamic roughness, while inland plains lack steep orographic features; both factors make ERA5 predictions more representative and less biased compared with complex terrain.

\subsection{Temporal Variability of Bias Correction Factors}
Based on the spatial distribution of the correction factors discussed in Section \ref{sec 3.2}, we turn to investigate their intra-annual dynamics at the monthly scale. Wind conditions and the resulting biases in reanalysis datasets are inherently variable across seasons and months, raising the question of whether the correction factors exhibit systematic temporal patterns. In this section, we examine their monthly distributions to assess whether the spatial regularities identified earlier remain consistent throughout the year, or whether deviations and extreme cases are seen during particular months. Such an analysis provides insights into the stability of the correction framework and reveals the bias patterns of reanalysis data across different time periods.

\begin{figure}[htbp]
    \centering
    \includegraphics[width=0.5\textwidth]{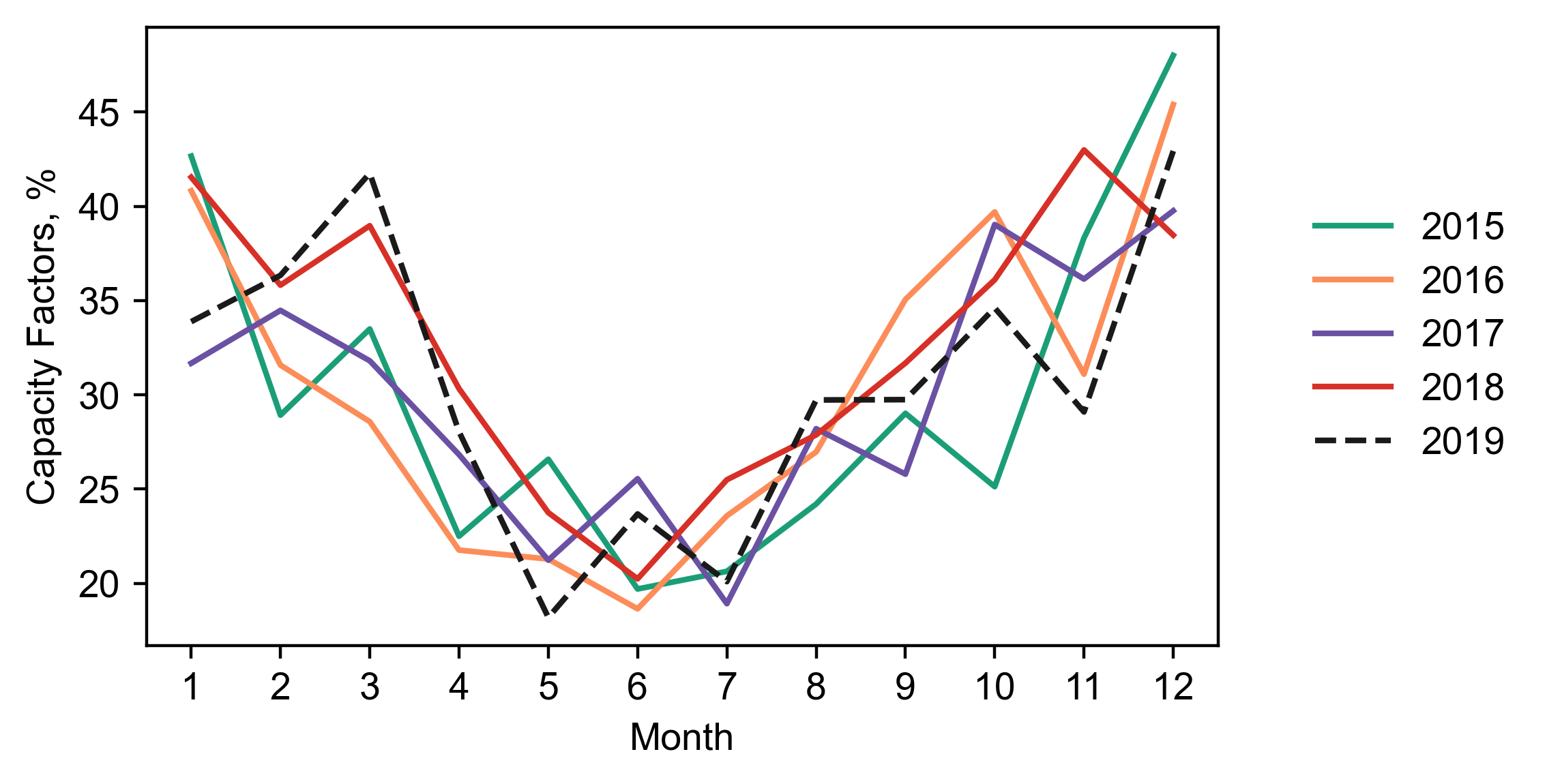}
    \caption{Monthly observed capacity factors in the UK, 2015–2019. Each line represents one year (training period: 2015–2018 shown in solid colours; test year 2019 shown as a dashed line).}
    \label{fig10}
\end{figure}

\autoref{fig10} presents the monthly capacity factors across the United Kingdom for 2015–2019. These values are derived from observed generation relative to installed capacity and therefore reflect the overall impact of variable wind resource conditions. By compiling and plotting these series, we can characterise the intra-annual evolution of wind power output, revealing the month-to-month variability in generation across the study period. \autoref{fig10} demonstrates a pronounced and recurrent intra-annual cycle in the observed capacity factors across the UK. The highest values are systematically recorded during the winter months (November–February), reflecting the dominance of stronger and more persistent wind resources during this period. In contrast, the lowest capacity factors occur during the warmer months of May–July, when calmer atmospheric conditions prevail and generation is substantially reduced. This seasonal cycle appears consistently across all years in the record, though the amplitude differs: 2015 and 2016, for instance, show particularly sharp contrasts between winter peaks and summer troughs, while 2017 exhibits a more muted range of variability. It is also noteworthy that the test year (2019) conforms to the same seasonal pattern without exhibiting any anomalous departures, thereby lending support to the reliability and stability of the modelling framework when applied outside the training period. The persistence of this winter–summer contrast highlights the strong influence of seasonal wind resource availability on national wind generation. Meanwhile, differences in the amplitude of this cycle across years reflect the influence of year-specific meteorological conditions, which can either amplify or dampen overall generation levels.
\begin{figure}[htbp]
    \centering
    \includegraphics[width=1\textwidth]{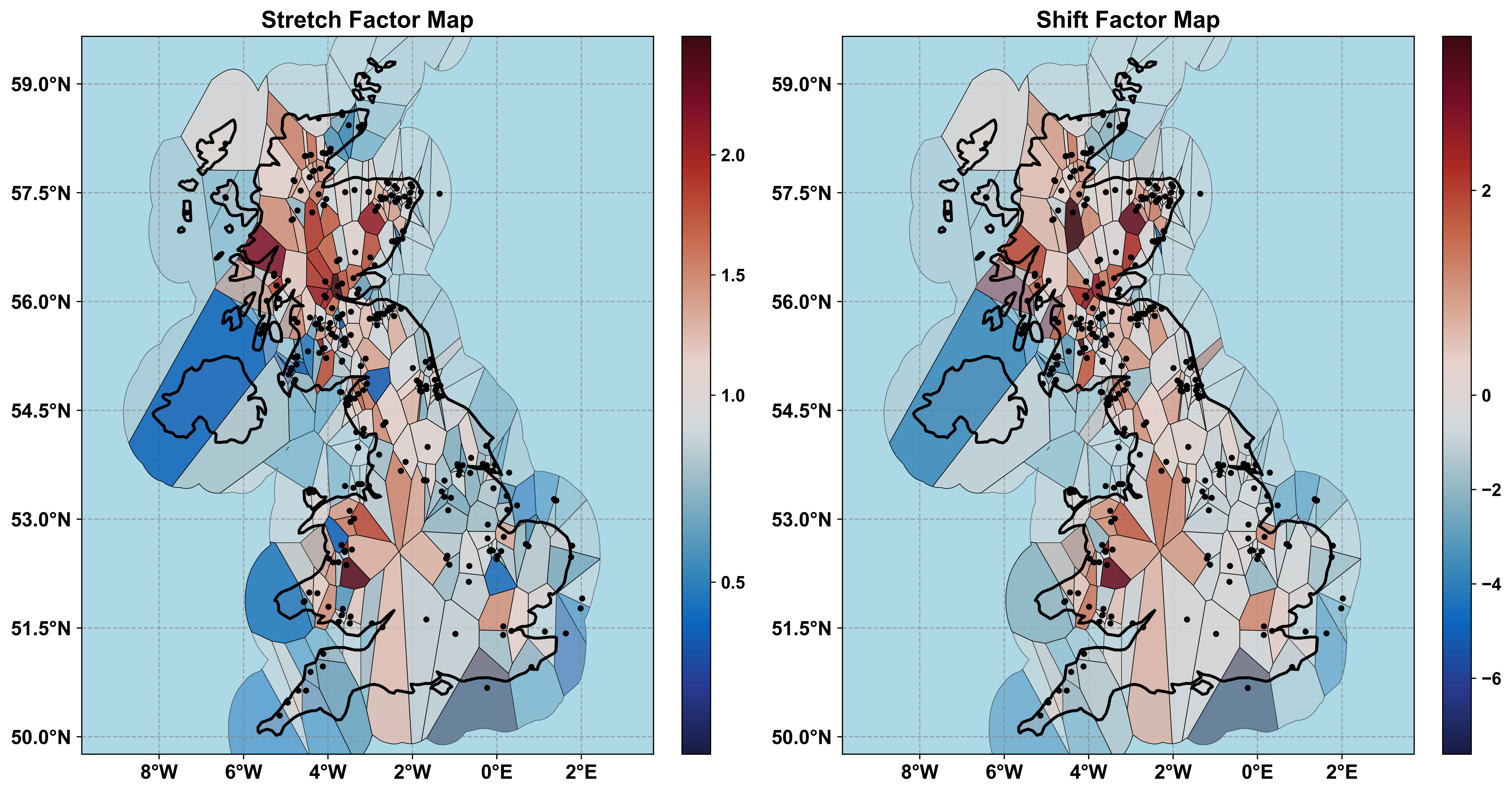}
    \caption{Bias correction factor distributions for June across the UK ($K=303$, monthly). (Left) Stretch factors ($\tilde{\alpha}$), and (right) shift factors ($\tilde{\beta}$). Red indicates higher correction values, and blue indicates lower values.}
    \label{fig11}
\end{figure}
\begin{figure}[htbp]
    \centering
    \includegraphics[width=1\textwidth]{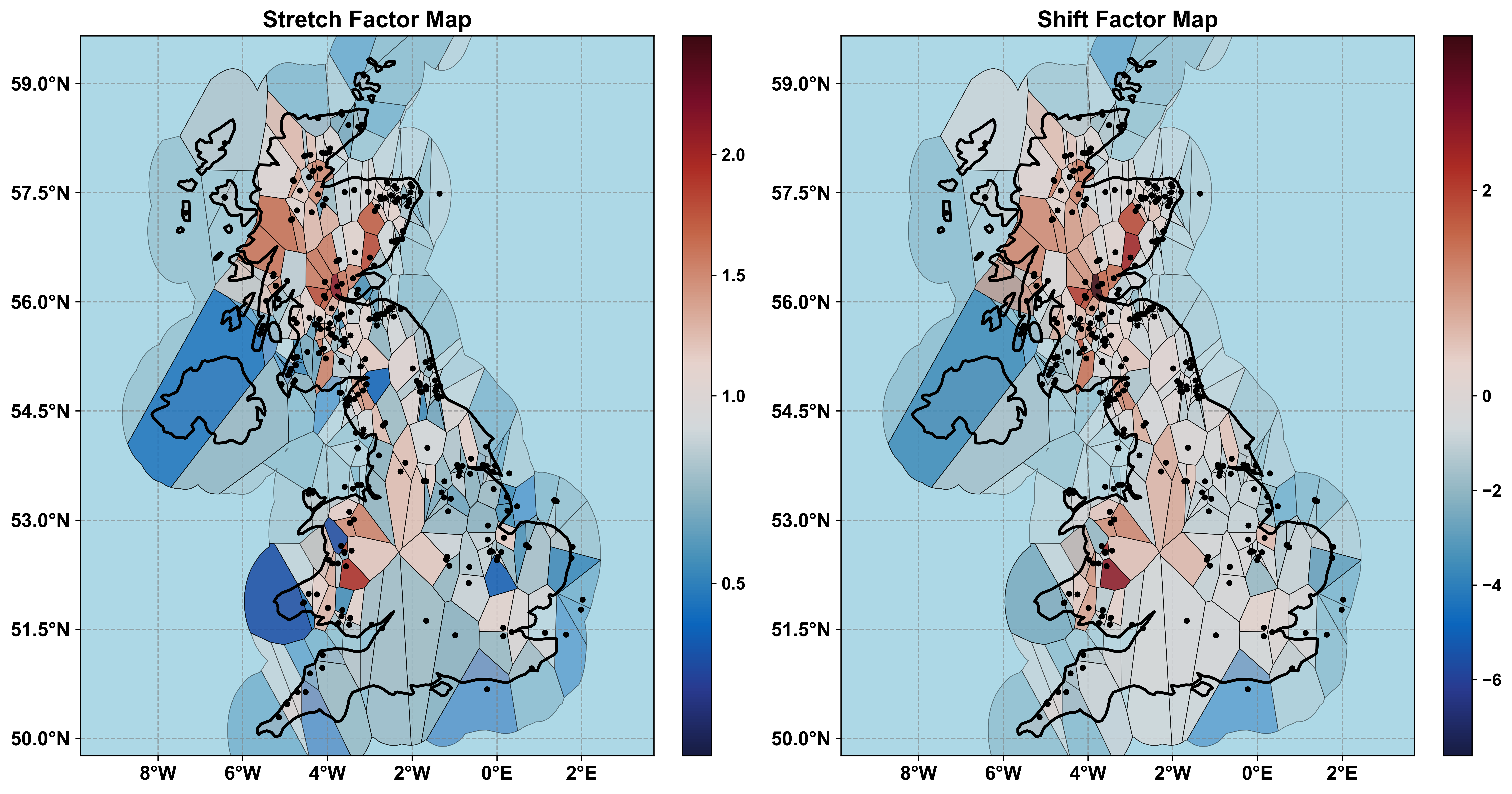}
    \caption{Bias correction factor distributions for December across the UK ($K=303$, monthly). (Left) Stretch factors ($\tilde{\alpha}$), and (right) shift factors ($\tilde{\beta}$). Red indicates higher correction values, and blue indicates lower values.}
    \label{fig12}
\end{figure}

Among the months of the year, June is characterized by relatively low levels of observed power generation, whereas December exhibits comparatively high levels. To illustrate these intra-annual contrasts, we select June and December as representative months for the subsequent analysis. As shown in \autoref{fig11}, the correction factors in June display a noticeable transition toward red tones across several regions, reflecting larger adjustment magnitudes relative to the annual mean distribution depicted in \autoref{fig8}. By contrast, December (\autoref{fig12}) exhibits the opposite tendency, with many regions undergoing a transition toward blue tones corresponding to smaller correction factors. This contrast reflects the underlying wind regime. Relative to the annual mean distribution, summer months with weaker wind speeds (e.g., June) require stronger upward corrections. Conversely, winter months with stronger winds (e.g., December) demand stronger downward corrections, as ERA5 more often overestimates wind speeds. The fundamental cause of this phenomenon is aligned with the bias mechanisms discussed earlier, but the deviations become amplified under extreme wind conditions: either very weak or very strong winds. Importantly, despite these monthly differences in magnitude, the overarching spatial distribution remains broadly robust, suggesting that the bias-correction framework captures structural spatial patterns that persist throughout the year.

\subsection{Correlation Analysis}
In the preceding discussion, we primarily examined the relationship between correction factors and elevation. Motivated by these insights, we extend our analysis to explore their associations with a broader set of geospatial features. Specifically, we introduce several variables (previously defined in Section \ref{Sec 2.6}), including elevation, hilliness, local turbine count, and distance to the sea. In this section, we present the distributions of these variables at the locations of the wind farms considered in this study. Subsequently, we conduct a correlation analysis to quantitatively assess their relationships with the correction factors, placing particular emphasis on the stretch factor $\tilde{\alpha}$.
\begin{figure}[htbp]
    \centering
    \includegraphics[width=1\textwidth]{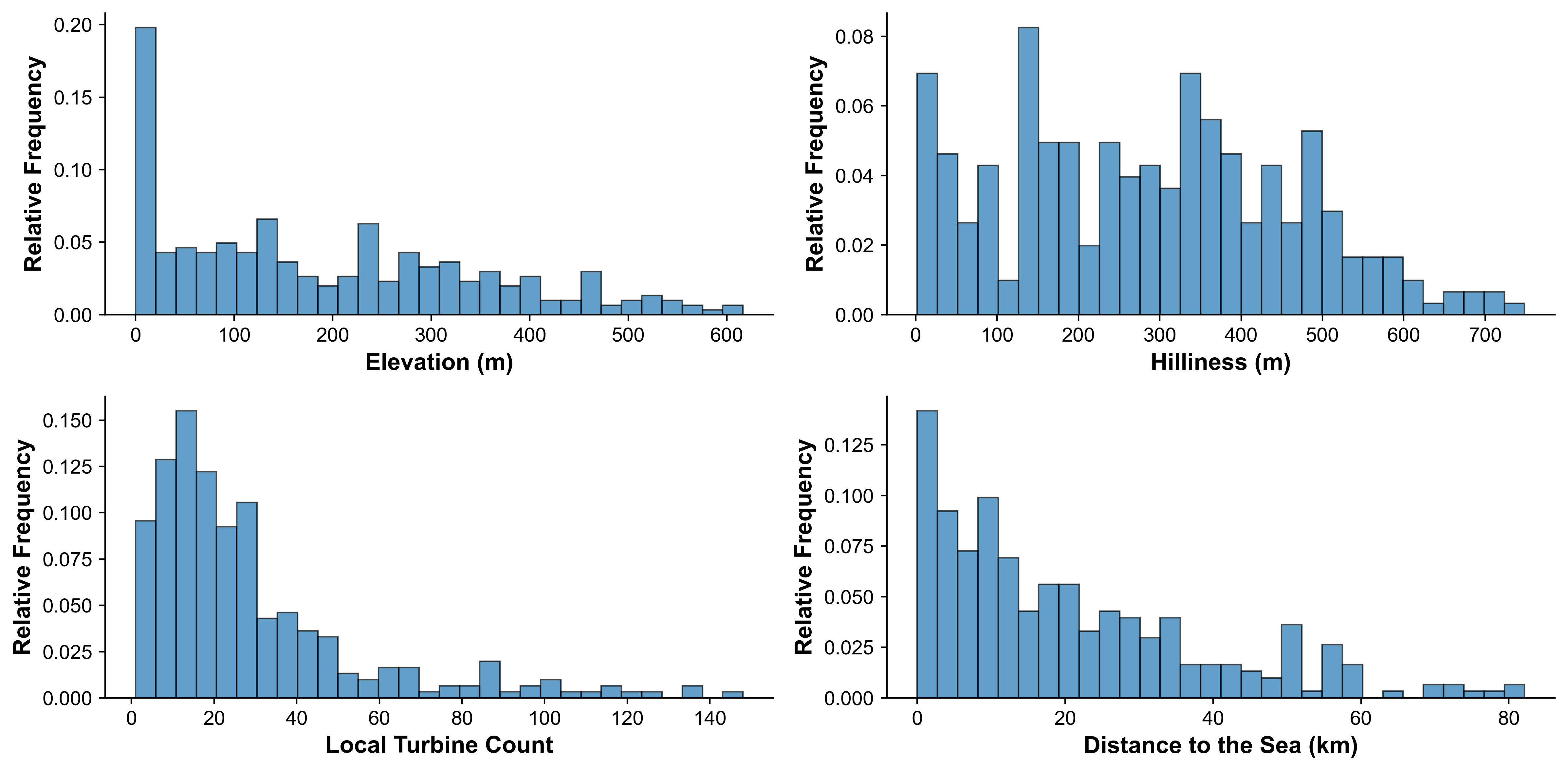}
    \caption{Histograms of geospatial characteristics: elevation, hilliness, local turbine amount, and distance to the sea.}
    \label{fig13}
\end{figure}

\autoref{fig13} presents the histograms of geospatial and turbine-related features at the locations of the wind farms considered in this study. The vertical axis represents the relative frequency, indicating the proportion of sites falling within each bin. As shown in \autoref{fig13}, the elevation of wind turbine sites in the UK is predominantly concentrated at low altitudes (below 200 m), while a considerable fraction is also found at intermediate elevations (200 – 450 m). Only a few sites reach higher elevations of up to 600 m, mainly in the Scottish Highlands and Welsh mountains. Hilliness does not simply reflect the absolute elevation of turbine sites, but rather captures the terrain variability within the surrounding 10 km radius, thereby providing a more representative measure of overall topographic complexity. The majority of sites are located in areas with hilliness below 400 m, while some locations exhibit terrain variability approaching 700 m. This distribution indicates that the UK landscape is characterised by extensive plains and lowlands, yet also contains a notable number of upland and mountainous areas with pronounced elevation differences, which are typically challenging for reanalysis datasets to resolve accurately. Beyond topography, local turbine count is introduced as an indicator of spatial concentration, serving as a coarse proxy for potential wake-related influence. This variable reflects the spatial concentration of turbines within a 7.5 km radius. Higher local turbine counts imply stronger wake interactions among neighbouring turbines, which in turn can lead to larger reductions in effective wind speed and increased discrepancies between simulated and observed generation. The variable distance to the sea is introduced to capture coastal influences on wind resources. Sites closer to the coastline are typically subject to stronger and less turbulent winds due to smoother surface roughness, whereas inland locations may experience reduced wind speeds and increased terrain-induced variability. It should be noted, however, that in reanalysis datasets coastal locations are sometimes overestimated because the limited spatial resolution blurs land–sea boundaries. The distribution of UK wind farms is strongly skewed toward short distances, with the majority of sites located within 25 km of the coastline. This strong skewness toward short coastal distances is partly intrinsic to the insular geography of the UK, where the absence of a large inland hinterland constrains wind farm siting and naturally biases the distribution toward coastal and offshore locations.
\begin{figure}[htbp]
    \centering
    \includegraphics[width=0.8\textwidth]{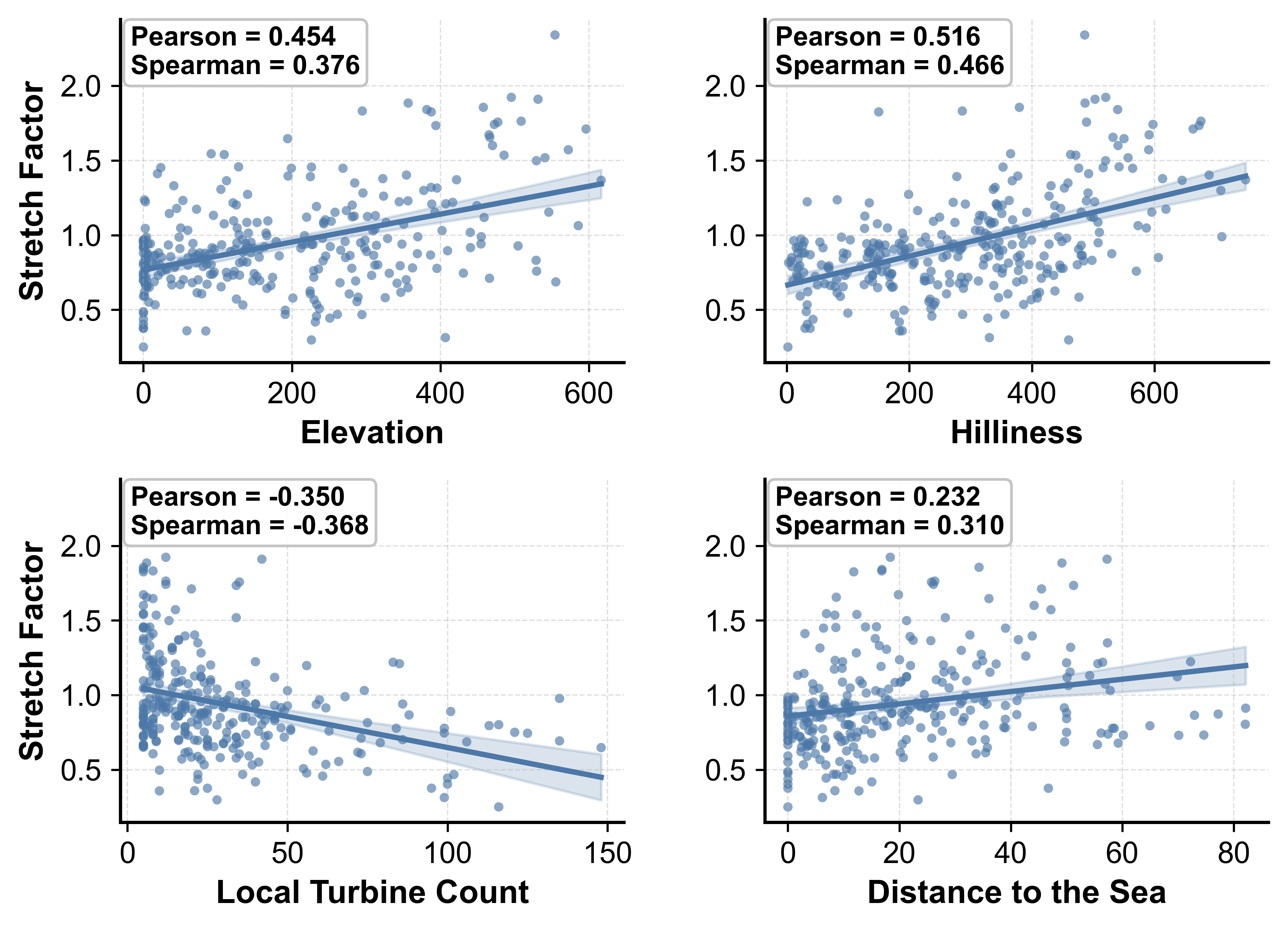}
    \caption{Scatter plots showing the correlation of the stretch factor ($\tilde{\alpha}$) with elevation, hilliness, local turbine count, and distance to the sea.  The Pearson and Spearman correlation coefficients \cite{schober2018correlation,deWinter2016} are reported in each panel.}
    \label{fig14}
\end{figure}

\autoref{fig14} and \autoref{tab:corr_geo} summarize the associations between the wind-speed bias correction factor (stretch factor $\tilde{\alpha}$) and selected geospatial features. Both Pearson’s $r$ and Spearman’s $\rho$ were computed to capture linear and monotonic relationships, respectively. In \autoref{tab:corr_geo}, all variables have \(p\)-values $<0.001$, indicating that the associations are statistically well supported. 
Moreover, the 95\% confidence intervals exclude zero, showing that the associations are unlikely to be spurious and are estimated with reasonable precision.

Among the four predictors, hilliness exhibits the strongest positive correlation with $\tilde{\alpha}$ ($r=0.52$, $\rho=0.47$), implying that sites located in more complex terrain require systematically stronger upward corrections. Hilly terrain induces substantial spatial variability in wind speeds, with local speed-ups and channeling effects along valleys and ridges. Turbines in such areas are typically sited using detailed on-site measurements to exploit these localized flow accelerations, leading to higher productivity relative to the surrounding terrain represented in the reanalysis data. Another strong positive correlation is observed with elevation ($r=0.45$, $\rho=0.38$), consistent with the known tendency of reanalysis products to underestimate wind speeds in higher-altitude orographically complex regions. The regression analysis shows that both elevation ($t=5.25$) and hilliness ($t=7.36$) remain significant when considered jointly. Their independent contributions suggest that elevation and local terrain complexity influence reanalysis biases through complementary mechanisms.

\begin{table}[htbp]
\centering
\caption{Correlation and regression analysis between the wind-speed bias correction factor (stretch factor $\tilde{\alpha}$) and geospatial features.
For Pearson’s $r$, $p$-values, and 95\% confidence intervals (CIs) are reported. 
For Spearman’s $\rho$, only the coefficients and $p$-values are shown.
The regression columns report the corresponding $t$-values and $p$-values.}
\label{tab:corr_geo}
\setlength{\tabcolsep}{6.3pt} 
\renewcommand{\arraystretch}{1.1} 
\begin{tabular}{lcccccccc}
\toprule
\multirow{2}{*}{\textbf{Variable}} & 
\multicolumn{4}{c}{\textbf{Pearson}} & 
\multicolumn{2}{c}{\textbf{Spearman}} &
\multicolumn{2}{c}{\textbf{Regression}} \\
\cmidrule(lr){2-5} \cmidrule(lr){6-7} \cmidrule(lr){8-9}
 & $r$  & $p$-value & $\mathrm{CI}_{\text{low}}$ & $\mathrm{CI}_{\text{high}}$ & $\rho$ & $p$-value & $t$-value & $p$-value \\
\midrule
Elevation        & 0.454   & $<0.001$ & 0.360 & 0.539 & 0.376 & $<0.001$ & 5.247  & $<0.001$  \\
Hilliness        & 0.516   & $<0.001$ & 0.428 & 0.594 & 0.466 & $<0.001$ & 7.362  &  $<0.001$ \\
Local Turbine Count  & -0.350  & $<0.001$ & -0.445 & -0.247 & -0.368 & $<0.001$ & -5.088  &  $<0.001$ \\
Distance to the Sea  & 0.232   & $<0.001$ & 0.123 & 0.336 & 0.310 & $<0.001$ &  3.419 & $<0.001$  \\
\bottomrule
\end{tabular}
\end{table}

By contrast, local turbine count shows a negative correlation ($r=-0.35$; 95\% CI $[-0.45,-0.25]$; $\rho=-0.37$) with the stretch factor. Because the VWF model used here (as described in Section \ref{sec:2}; \cite{staffell2016using,staffell2016renewables,benmoufok2024improving}) does not explicitly represent intra-farm wake losses, we include local turbine count as an explanatory variable intended to partially capture wake-related effects at aggregated scales. While this is not a substitute for a dedicated wake model, it may help explain the biases arising when site-level winds deviate from the reanalysis background flow due to turbine interactions. Denser layouts and clusters of neighbouring farms increase cumulative wake losses and array blockage that are not resolved at the reanalysis grid scale, which depress local winds relative to the background flow. As a result, the site-level winds tend to be lower than the reanalysis would suggest, and the required bias-correction factor is smaller in high-density areas. The consistency between Pearson and Spearman estimates indicates that the relationship is monotonic and not driven by a few outliers. In addition, the multiple regression result ($t=-5.09$) highlights its importance as an independent predictor. These consistent correlations demonstrate the role of local turbine count as an indicator implicitly capturing wake effects and suggest that future work could improve the VWF model by explicitly incorporating these effects.

Finally, distance to the sea exhibits a weaker but still positive correlation. Although the effect size is modest, the narrow 95\% confidence interval $[0.12, 0.34]$ confirms that the relationship is robust. In particular, we observe that Spearman’s $\rho$ exceeds Pearson’s $r$ (with $\rho>0.3$), indicating that while the linear association is not especially strong, the monotonic trend remains clearly evident. Moreover, the multiple regression result ($t = 3.42$) confirms its non-negligible independent contribution. This reflects the fact that turbines located closer to the coastline and those situated further inland are subject to distinct wind regimes shaped by their geographic environments.

Overall, the correlation analysis reveals consistent and interpretable patterns linking the wind-speed bias correction factor to geospatial features. These associations show that both orographic complexity and anthropogenic modifications of the flow environment exert systematic influences on reanalysis biases, whereas coastal proximity plays a secondary yet detectable role.

\section{Conclusions}
\label{sec:4}
This study applies, extends, and analyses a high-resolution, cluster-based bias correction framework for reanalysis wind speeds that increases agreement with observed wind power performance. In addition, it systematically identifies and analyses the terrain-induced geographic causes of these biases, associating error structures with orographic complexity and coastal proximity. By integrating spatially resolved corrections with detailed geographic analysis across wind farm sites, the framework demonstrates robustness across two state-of-the-art reanalysis datasets (ERA5 and MERRA-2) and achieves consistent improvements in wind simulation accuracy. These advances yield a deeper understanding of error patterns in the United Kingdom.

The effectiveness of this framework is further demonstrated by quantitative evaluation. Results show that the bias correction reduces ERA5-based wind power simulation errors by more than 32\% when 303 clusters are applied. MERRA-2 shows a comparable effect, with RMSE reduced by 34\%, which illustrates that the bias correction framework is robustly applicable across different reanalysis products. In addition, the refined model reduces both random errors (RMSE and MAE) and systematic deviations, with mean bias errors shifting closer to zero after correction. Collectively, these findings indicate that the proposed approach improves predictive accuracy and provides insights into the spatial structure of reanalysis wind errors at a finer granularity than conventional methods.

In addition to these quantitative gains, the spatial bias mapping revealed distinct patterns in reanalysis discrepancies. ERA5 wind speeds were systematically underestimated in high-elevation, complex terrain, particularly in the Scottish Highlands and mountainous regions of Wales, where stretch correction factors $\tilde{\alpha}$ exceeded 1. In contrast, overestimation occurred in certain low-lying coastal areas of England, with correction factors often between 0.5 and 0.8, indicating a need for downward adjustment. Most flat inland plains and offshore sites exhibited relatively minor, consistent biases ($\tilde{\alpha}$ typically within 0.8--1.2), suggesting that reanalysis data are more reliable over homogeneous terrain and marine environments. To complement these spatial patterns, correlation analysis further showed that elevation, hilliness, and distance to the sea are positively associated with the magnitude of correction factors, whereas local turbine count exhibited a negative correlation. In summary, these spatial patterns indicate that topography, the spatial concentration of turbines, and land-sea contrasts are key factors shaping reanalysis wind errors. By systematically mapping these biases across the United Kingdom, this study fills an important gap in the literature and demonstrates the necessity of accounting for geographic heterogeneity in bias correction, since spatially uniform correction approaches risk neglecting regional error characteristics.

Within the broader renewable energy transition, this work contributes to more reliable wind resource assessment and power system planning. By improving the fidelity of site-specific wind and power simulations, the framework can help better inform siting and investment decisions and reduce uncertainty in expected yields. The spatial diagnosis of reanalysis biases also provides empirically derived inputs to guide climate model calibration and to support downscaling methodologies, including both statistical and dynamical approaches, thereby promoting greater consistency between modelled wind fields and observations. Overall, these contributions help mitigate modelling uncertainty and offer a practical basis for system-integration studies, operational planning, and decision making.

\section*{CRediT authorship contribution statement}
\textbf{Yan Wang:} Writing – original draft, Writing – review \& editing, Conceptualization, Formal analysis, Data curation, Software, Investigation, Visualization. \textbf{Simon C. Warder:} Conceptualization, Writing – review \& editing, Validation. \textbf{Ellyess F. Benmoufok:} Software. \textbf{Andrew Wynn:} Writing – review \& editing, Conceptualization, Supervision. \textbf{Oliver R.H. Buxton:} Writing – review \& editing, Conceptualization, Supervision. \textbf{Iain Staffell:} Resources, Methodology. \textbf{Matthew D. Piggott:} Supervision, Methodology, Project administration, Writing – review \& editing.

\section*{Declaration of competing interest}
The authors declare that they have no known competing financial interests or personal relationships that could have appeared to influence the work reported in this paper.

\section*{Acknowledgements}
We acknowledge the funding provided by the Taylor donation to Imperial College London, via the Grantham Institute and the Energy Futures Lab. Yan Wang acknowledges additional support from the Department of Aeronautics and the Department of Earth Science and Engineering at Imperial College London.

\section*{Data availability}
Data will be made available on request. 

\bibliography{references}

\end{document}